\documentclass[a4paper,11pt]{article}
\usepackage{jcappub} 

\usepackage{xcolor}       
\usepackage{bm}           
\usepackage{multirow}
\usepackage{array}
\usepackage{comment}
\usepackage[flushleft]{threeparttable}

\title{One-loop kernels in scale-dependent Horndeski theory}

\author[a]{Ziyang Zheng,}
\author[a]{Hanqiong Jia,}
\author[a]{Bilal T\"udes,}
\author[b]{Anton Chudaykin,}
\author[b]{Martin Kunz,}
\author[a,c]{Luca Amendola}

\affiliation[a]{Institut f\"ur Theoretische Physik, Ruprecht-Karls-Universit\"at Heidelberg,\\
Philosophenweg 16, D-69120 Heidelberg, Germany}
\affiliation[b]{D\'epartement de Physique Th\'eorique and Center for Astroparticle Physics,\\
Universit\'e de Gen\`eve, Quai E. Ansermet 24, CH-1211 Gen\`eve 4, Switzerland}
\affiliation[c]{New York University Abu Dhabi, PO Box 129188, Abu Dhabi, United Arab Emirates and Center for Astrophysics and Space Science (CASS), New York University Abu Dhabi}

\emailAdd{Zheng@thphys.uni-heidelberg.de}
\emailAdd{Jia@thphys.uni-heidelberg.de}
\emailAdd{tuedes@thphys.uni-heidelberg.de}
\emailAdd{anton.chudaykin@unige.ch}
\emailAdd{martin.kunz@unige.ch}
\emailAdd{l.amendola@thphys.uni-heidelberg.de}

\abstract{
We investigate the nonlinear evolution of cosmological perturbations in theories with scale-dependent perturbation growth, first in general and then focusing on Horndeski gravity. Within the framework of standard perturbation theory, we derive the second- and third-order kernels and show that they are fully determined by two effective functions, \( h_1 \) and \( h_c \), which parametrize deviations from general relativity. Using the Wronskian method, we obtain solutions for the nonlinear growth functions and present explicit expressions for the resulting kernels, including bias and redshift space distortions, valid in the limit in which the $k$-dependent part is subdominant. We show that the kernels are entirely dependent on the linear growing mode: once this is calculated, the kernels are analytic up to a time integral. We also include redshift-space distortions (RSD) and scale-dependent bias. Our approach provides a physically motivated framework for evaluating the one-loop galaxy power spectrum in scale-dependent theories, suitable for the forecasts and actual data analysis.
}

\begin{document}
\maketitle
\flushbottom


\section{Introduction}\label{sec:intro}

The large-scale structure (LSS) of the Universe firmly establishes itself as a reliable probe of cosmology and fundamental physics. 
This provides important complementary information to that from the cosmic microwave background radiation (CMB) (e.g. {\it Planck}~\cite{Planck:2018vyg}, ACT~\cite{ACT:2020gnv}, SPT-3G~\cite{SPT-3G:2021eoc}) and supernova (SN Ia) measurements. 
Cosmic surveys such as BOSS~\cite{BOSS:2016wmc}, eBOSS~\cite{eBOSS:2020yzd} and the two-year data releases from DESI~\cite{DESI:2025zgx} have played an important role in constraining cosmological parameters and testing various theoretical models.
Forthcoming Stage-IV experiments -- DESI, Euclid~\cite{Euclid:2024yrr}, and the Vera C. Rubin Observatory~\cite{LSST:2008ijt} -- are expected to significantly increase the amount of cosmological information, potentially reaching sub-percent precision in parameter constraints. 
Such a level of precision in observational data requires an accurate theoretical description of galaxy clustering. 

In recent years, the full-shape analysis has become a standard method for extracting cosmological information from spectroscopic surveys. This approach builds on the Effective Field Theory (EFT) of LSS~\cite{Baumann:2010tm,Carrasco:2012cv}, which provides an accurate and mathematically consistent theoretical framework for the clustering of matter and various luminous tracers in the mildly nonlinear (quasilinear) regime. 
The idea of this approach is to model the full-shape power spectrum directly and place constraints on the model parameters.
This is akin to the analysis of CMB data, and enhances the cosmological utility of current and upcoming surveys.
Importantly, the full-shape analysis models the broadband shape of the galaxy power spectrum, and hence extract the information which is not accessible with the conventional BAO/RSD techniques.
The EFT-based approach has been successfully applied to the BOSS galaxy samples in the context of $\Lambda$CDM~\cite{Ivanov:2019pdj,DAmico:2019fhj,Philcox:2021kcw}, dynamical dark energy~\cite{DAmico:2020kxu,Chudaykin:2020ghx}, early dark energy~\cite{Ivanov:2020ril,Chudaykin:2020igl}, ultra-light axion dark matter~\cite{Lague:2021frh,Rogers:2023ezo}, and model-independent analysis \cite{Schirra:2024rjq}.

The LSS data can be also used to test gravity. General Relativity (GR) has been validated on planetary scales using Parameterized Post Newtonian (PPN) parameters~\cite{Will:2014kxa}.
Most previous tests of gravity on cosmological scales~\cite{Bellini:2015xja,Noller:2018wyv} relied on the traditional RSD analysis, which measure the amplitude of fluctuations $f\sigma_8$.
However, these analyses assume a fixed-shape template for the linear matter power spectrum computed within the $\Lambda$CDM model, and therefore such tests of modified gravity are neither self-consistent nor model independent. More general, model-independent tests of gravity have been conducted, obtaining robust but, so far, weak constraints \cite{Pinho:2018unz, Zheng:2023yco, Sakr:2025xem}.
In contrast, the EFT-based full-shape analysis recalculates the shape of the linear matter power spectrum as a function of cosmological parameters, offering a consistent framework for testing gravity.
In addition, incorporating the one-loop correction allows the inclusion of a larger number of Fourier modes, increasing the useful cosmological information, tightening parameter constraints, and breaking degeneracies. 

Nonlinear corrections to the power spectrum in scale-dependent modified gravity theories are computationally expensive, as the analytic form of the perturbative kernels are unknown, except in some simple cases \cite{Tudes:2024jpg}. 
This requires solving differential equations that depend not only on the wavenumber configuration but also on the cosmological parameters. Due to this complexity, performing a MCMC analysis for parameter estimation becomes hardly feasible.
Most full-shape analyses in the context of modified gravity therefore rely on the commonly used Einstein–de Sitter (EdS) kernels~\cite{Ishak:2024jhs,Chudaykin:2024gol}.~\footnote{Refs.~\cite{Piga:2022mge,Taule:2024bot} utilize the more general bootstrap approach of~\cite{DAmico:2021rdb}, which is applicable to generic scale-independent modified gravity models.}
This approach is valid when the growth rate and matter density parameter satisfy $f^2 = \Omega_m$. 
However, this relation can fail if the background dynamics deviates from the $\Lambda$CDM model, as in the normal branch of DGP gravity~\cite{Piga:2022mge}, or in scale-dependent modified gravity scenarios~\cite{Amendola:2003wa,Koyama:2009me,Aviles:2020wme} (where the linear growth function exhibits both time and scale dependence). Early efforts to explore the power spectrum in general classes of modified gravity and redshift-space distortions (RSD) are presented in \cite{Taruya_2016,Bose_2016}. A more recent and comprehensive analysis of the uncertainties arising from various modeling assumptions and approximations in the nonlinear redshift-space galaxy power spectrum for cosmologies beyond $\Lambda$CDM can be found in \cite{EuclidP2024}. A recent study~\cite{Aviles:2024zlw} presented constraints on the $f(R)$ model using the power spectrum within the one-loop EFT framework.

In this work, we extend the Wronskian method, earlier adopted in the non-linear perturbation context in \cite{Takushima:2015iha, Bartolo:2013ws}, to provide analytic expressions (up to a time integral) for the second- and third-order kernels in general models with scale-dependent perturbation growth.
We reformulate the approach of~\cite{Aviles:2021que,Rodriguez-Meza:2023rga}, developed within the Lagrangian Perturbation Theory, in the Eulerian framework. By employing the Wronskian method, we derive solutions for the nonlinear growth functions. We then generalize the scale-dependent kernels for bias tracers in redshift space. Our approach is applicable to any cosmological scenario that involves an additional massive degree of freedom, such as Horndeski gravity. In particular, it can be used to compute accurate perturbative kernels in the presence of massive neutrinos, where scale dependence is introduced through free-streaming.~\footnote{The effect of massive neutrinos can be modeled within the single-fluid approach by using the linear neutrino transfer function in the Poisson equation~\cite{Lesgourgues:2009am}.} 

Our paper is structured as follows.  In Sec.~\ref{sec:Linear}, we provide the general formalism for scale-dependent linear growth and derive expressions for the linear growth function and growth rate. In Sec.~\ref{sec: kernels}, we present the derivation of the second- and third-order standard perturbation theory (SPT) kernels, valid in any theory with scale-dependent perturbation growth. In Sec.~\ref{sec:F2G2F3G3}, we specialize these results to Horndeski gravity, expressing the nonlinear kernels in terms of two effective functions, \( h_1 \) and \( h_c \). In Sec.~\ref{sec: bias_RSD}, we include galaxy bias and redshift-space distortions, and in Sec.~\ref{sec: Pk at oneloop} we assemble all ingredients into the one-loop galaxy power spectrum. We conclude in Sec.~\ref{sec:con}. Technical details and derivations are collected in the Appendices.


\section{Linear equations}

\label{sec:Linear}

We begin by considering the general evolution equation for the linear
matter growth function $D$,
\begin{align}
D''+\mathcal{F}D'-S\,D & =0\,.\label{eq:Dz-1-1}
\end{align}
We choose units such that $8\pi G=M_{p}^{-2}=1$, where $M_{p}$ denotes the reduced Planck mass, and a prime denotes differentiation with
respect to the e-folding time $N=\ln a$, where $a$ is the scale
factor. The function $\mathcal{F}(N)$ represents a generalized, time-dependent
friction term. In the standard case, it is given by $\mathcal{F}=2+H'/H$,
but additional terms may arise, for instance, when the equivalence
principle is violated (cf. Refs.~\cite{Amendola:2001rc,Castello:2024lhl})
or in  presence of viscous dark matter (cf. Refs.~\cite{Barbosa:2018iiq,Velten:2013pra}).
While we treat $\mathcal{F}$ as a purely time-dependent function
in this work, we note that it could, in principle, exhibit scale dependence.
Similarly, for the  term $S$ (referred to as "source" term since it comes from the right-hand-side of the Poisson equation), the standard expression $S=3\Omega_{m}(N)/2$,
where $\Omega_{m}(N)$ denotes the time-dependent matter density parameter,
may receive corrections in scenarios involving modified gravity (see,
e.g., Refs.~\cite{Amendola_2013,Planck:2018vyg,Euclid:2023rjj,Joudaki:2020shz})
or the presence of massive neutrinos. For the explicit form in the
latter case, we refer the reader to Eq. (2.2)
in Ref.~\cite{Aviles:2021que}.
In this paper, we assume that any such correction is small with respect
to the standard part and we expand systematically our expressions to first order in the correction.

We now decompose explicitly the linear growth function $D$ and the
source term $S$ into a purely time-dependent part (subscript $z$)
and a sub-dominant scale-dependent correction (subscript $kz$): $D=D_{z}+\varepsilon D_{kz}$
and $S=S_{z}+\varepsilon S_{kz}$, where we use the order parameter
$\varepsilon$ to keep track of the sub-dominant terms. At zero-th
order in $\varepsilon$ we have 
\begin{align}
D_{z}''+\mathcal{F}D_{z}'-S_{z}D_{z} & =0\,.\label{eq:Dz-1-2}
\end{align}
We denote the solutions of this equation by $D_{\pm}$. At first order
we obtain instead
\begin{align}
    D_{kz}''+\mathcal{F}D_{kz}'-S_{z}D_{kz} & =S_{kz}D_{z}\,.\label{eq:dkz-1}
\end{align}
Once a solution $D_{+}$ is known (assumed to be the fastest growing
mode), numerically or analytically, the decaying mode $D_{-}$ can
be determined via 
\begin{equation}
D_{-}=D_{+}\left[\int_{N_{0}}^{N}\frac{e^{-\int_{N_{0}}^{x}\mathcal{F}(\bar{x})\,\bar{x}}}{D_{+}^{2}}\,\mathrm{d}x+C\right],
\end{equation}
where $C$ is a constant determined by the initial condition of
$D_{-}$ at $N=N_{0}$. In general, therefore, $D_{z}$ is a linear
combination of $D_{\pm}$ and we assume as usual that
only $D_{+}$ survives  at late times.

Once we know $D_{\pm}$, we can solve Eq.~\eqref{eq:dkz-1} for the
scale-dependent correction $D_{kz}$ using the Wronskian method: 
\begin{equation}
D_{kz}=- D_{+}\int_{N_{0}}^{N}\mathrm{d}x\frac{D_{-}D_{+} S_{kz}}{W}+ D_{-}\int_{N_{0}}^{N}\mathrm{d}x\frac{D_{+}^{2} S_{kz}}{W}
\label{eq: D_kz_withW}
\end{equation}
where (Abel's formula) 
\begin{equation}
W(D_{+},D_{-})=D_{+}D_{-}'-D_{+}'D_{-}=W_{0}\exp\Big[-\int_{N_{0}}^{N}\mathcal{F}(x)\mathrm{d}x\Big]\label{eq:Wronskian_2-1}
\end{equation}
is the Wronskian of the homogeneous solution and $W_{0}$ is its value
at $N=N_{0}$. Since we only need a particular solution of the inhomogeneous
equation (\ref{eq:dkz-1}), we can take $W_{0}=1$. Note that the normalization
of $D_{-}$ is irrelevant, as any prefactor cancels between $D_{-}$
and the Wronskian $W(D_{+},D_{-})$ in Eq.~\eqref{eq:Wronskian_2-1},
whereas the normalization of $D_{+}$ enters the particular solution
of $D_{kz}$ and must be chosen consistently.

The growth rate $f$ is defined as $f=D'/D$, and as above we can
define a $k$-independent part $f_{z}=D_{z}'/D_{z}$ at zero-th order
in $\varepsilon$, that obeys the equation
\begin{equation}
f_{z}'+f_z^2+\mathcal{F}f_{z}-S_{z}=0\, ,\label{eq:evolution_f_z}
\end{equation}   
and a first order, $k$-dependent part $f_{kz}$ that obeys the
equation
\begin{equation}
f_{kz}'+f_{kz}(\mathcal{F}+2f_{z})-S_{kz}=0\,.\label{eq:evolution_f_k}
\end{equation}
The solution to Eq.~\eqref{eq:evolution_f_k} is 
\begin{equation}
f_{kz}(k,N)=c_{1}e^{-I_{1}(N)}+e^{-I_{1}(N)}\int_{N_{0}}^{N}e^{I_{1}(x)}S_{kz}(k,x)\mathrm{d}x\,,\label{eq:fk_solution_full}
\end{equation}
where $I_{1}(N)=\int_{N_{0}}^{N}(2f_{z}+\mathcal{F})\mathrm{d}x$.
Imposing the boundary condition that scale-dependence is negligible at high redshifts, $f_{kz}(N_{0})=0$ as $N_{0}\to-\infty$,
we set $c_{1}=0$, yielding 
\begin{equation}
f_{kz}(k,N)=e^{-I_{1}(N)}\int_{N_{0}}^{N}e^{I_{1}(x)}S_{kz}(k,x)\mathrm{d}x \,.
\label{eq:fk_solution}
\end{equation}
The full linear growth rate at a given scale $k$ is therefore given
by the sum of the scale-independent part $f_{z}$ and the scale-dependent
correction $f_{kz}$.

Since later on we will focus on Horndeski's model, we discuss now
this case. Within the Horndeski framework, the source term takes the
form (see e.g. Ref.~\cite{Amendola:2019laa}) 
\begin{align}
S(k,N)\equiv\frac{3}{2}\Omega_{m}(N)h_{1}\left(\frac{1+h_{5}k^{2}}{1+h_{3}k^{2}}\right) & =S_{z}+S_{kz}\,,\label{eq: A_k_honedeski}
\end{align}
where the $h_i$ are functions of time only, and where
we defined 
\begin{equation}
\label{sources}
S_{z}\equiv\frac{3}{2}\Omega_{m}(N)h_{1}\,,\quad S_{kz}\equiv\frac{3}{2}\Omega_{m}(N)h_{1}\frac{(h_{5}-h_{3})k^{2}}{1+h_{3}k^{2}}\equiv\frac{3}{2}\Omega_{m}(N)h_{c}(k,N)\,,
\end{equation}
with $h_{c}\equiv h_{1}(h_{5}-h_{3})k^{2}/(1+h_{3}k^{2})$. As already
mentioned, we will always assume that the $k$- dependent correction
is sub-dominant; this means we treat $h_{c}$ as our order parameter
(and therefore we do not need any longer the parameter $\varepsilon$).
We see that only two effective Horndeski functions can be constrained:
$h_{1}$, which depends only on time, and $h_{c}$, which depends
on both time and scale. Their relation with the alternative $\alpha$-parametrization
is discussed in App. \ref{sec:alpha}. To illustrate the effects of scale-dependent gravity, we also compare the linear growth rate \( f \) in Horndeski gravity with their \(\Lambda\)CDM counterparts in App.~\ref{sec:numerical_test}.

In summary, the scale-dependent linear growth factor $D(\mathbf{k},N)$
and the growth rate $f(\mathbf{k},N)$ are fully characterized by
the Horndeski parameters $h_{1}$ and $h_{c}$, via the linear growth
$D_{+}$.


\section{General kernels of standard perturbation theory} \label{sec: kernels} %

In this section, we briefly review the derivation of the second-order SPT kernels in Sec.~\ref{sec:2nd_FG}, following the standard method (see, e.g. App.~A of Ref.~\cite{Aviles:2021que}), and extend the formalism to third order in Sec.~\ref{sec:3rd_FG}.
The kernels obtained in this section are completely general and can be applied to any scale-dependent growth. In Sec. \ref{sec:F2G2F3G3} we specialize to Horndeski and take the first order limit in $S_{kz}$. %

The evolution of the density contrast $\delta$ and the velocity divergence $\theta$ is governed by the continuity and Euler equations. In Fourier space, they are given by \footnote{
The integrals are defined as 
\begin{equation}
    \int_{\sum\mathbf{k}_{i}=\mathbf{k}}[...]=\int [\prod _i\frac{\mathrm{d}^3\mathbf{k}_i}{(2\pi)^3}] (2\pi)^3\delta_D \left( \sum_i\mathbf{k}_i-\mathbf{k} \right) [...]  \nonumber \,.
\end{equation}
Throughout this work, we adopt the Fourier transform convention
\begin{equation}
\tilde{f}(\mathbf{k}) = \int \mathrm{d}^3\mathbf{x} \, f(\mathbf{x})\, e^{-i \mathbf{k} \cdot \mathbf{x}}\,, \qquad
f(\mathbf{x}) = \int \frac{\mathrm{d}^3\mathbf{k}}{(2\pi)^3} \, \tilde{f}(\mathbf{k})\, e^{i \mathbf{k} \cdot \mathbf{x}} \nonumber \,,
\end{equation} 
such that the Dirac delta function satisfies
\begin{equation}
(2\pi)^3 \delta_D(\mathbf{k}) = \int \mathrm{d}^3\mathbf{x} \, e^{i \mathbf{k} \cdot \mathbf{x}} \nonumber \,.
\end{equation} 
}: %
\begin{align}
\delta'_{\mathbf{k}} - \theta_{\mathbf{k}} &= \int_{\mathbf{k}_{12} = \mathbf{k}} \alpha_{1,2}\, \theta_{\mathbf{k}_1}\, \delta_{\mathbf{k}_2}\,, \label{eq:continuity}
\\
\theta'_{\mathbf{k}} + \mathcal{F}\, \theta_{\mathbf{k}} - S(k)\,\delta_{\mathbf{k}} &= \int_{\mathbf{k}_{12} = \mathbf{k}} \beta_{1,2}\, \theta_{\mathbf{k}_1}\, \theta_{\mathbf{k}_2}\,,
\label{eq:Euler}
\end{align}
where  we adopt the shorthand notation \( 
\theta_{\textbf{k}} = \theta(\textbf{k}), \delta_{\textbf{k}} = \delta(\textbf{k}) \), \( \theta \equiv -i k_i v^i / (a H ) \) is the rescaled velocity divergence and \( \mathbf{k}_{ij}=\mathbf{k}_i+\mathbf{k}_j \), and where
 the mode-coupling functions $\alpha$ and $\beta$ are given by 
\begin{equation}
\alpha_{1,2} = 1 + \frac{\mathbf{k}_1 \cdot \mathbf{k}_2}{k_1^2}\,, \qquad 
\beta_{1,2} = \frac{k_{12}^2\, (\mathbf{k}_1 \cdot \mathbf{k}_2)}{2 k_1^2 k_2^2}\,.
\end{equation}

In SPT, the nonlinear evolution of $\delta$ and $\theta$ is captured by expanding them order-by-order in powers of the linear density field, $\delta(\textbf{k},N)=\sum_{n=1}^\infty\delta^{(n)}(\textbf{k},N)$ and $\theta(\textbf{k},N)=\sum_{n=1}^\infty\theta^{(n)}(\textbf{k},N)$. At $n$-th order, the solutions are written as convolutions of time- and scale-dependent SPT kernels $F_n$ and $G_n$ with $n$ copies of the linear field:
\begin{equation}
\begin{aligned} \label{eq: kernels}
\delta^{(n)}(\mathbf{k}, N) &= \int_{\mathbf{k}_1 + \cdots + \mathbf{k}_n = \mathbf{k}} F_n(\mathbf{k}_1, \ldots, \mathbf{k}_n; N)\, \delta^{(1)}_{\mathbf{k}_1} \cdots \delta^{(1)}_{\mathbf{k}_n}\,, \\
\theta^{(n)}(\mathbf{k}, N) &= \int_{\mathbf{k}_1 + \cdots + \mathbf{k}_n = \mathbf{k}} G_n(\mathbf{k}_1, \ldots, \mathbf{k}_n; N)\, \delta^{(1)}_{\mathbf{k}_1} \cdots \delta^{(1)}_{\mathbf{k}_n}\,.
\end{aligned}
\end{equation}
The kernels $F_n$ and $G_n$ encode the nonlinear mode coupling generated by gravitational evolution. It is straightforward to verify that
\begin{equation} \label{eq:F_1_G_1}
   F_1 = 1  ;\quad G_{1}=f(k) \,.
\end{equation}

The standard Poisson equations for the potential $\phi$ that we adopted to derive Eqs. (\ref{eq:continuity}-\ref{eq:Euler}) should be modified in Horndeski to take into account the Vainshtein effect in the perturbation expansion. The resulting corrected Poisson equation in Fourier space up to third order has been derived in \cite{Cusin_2018} in the $k$-independent limit:
\begin{align}
    -\frac{k^{2}}{\mathcal{H}^{2}}\phi&=S\delta+S_{2}\int_{\mathbf{k}_{12}=\mathbf{k}}\delta^{(1)}_{\mathbf{k}_{1}}\delta^{(1)}_{\mathbf{k}_{2}}\gamma_{2}(\mathbf{k}_{1},\mathbf{k}_{2}) \label{eq: high_Poisson_1}
    \\&+S_{22}\int_{\mathbf{k}_{123}=\mathbf{k}}\delta^{(1)}_{\mathbf{k}_{1}}\delta^{(1)}_{\mathbf{k}_{2}}\delta^{(1)}_{\mathbf{k}_{3}}\gamma_{2}(\mathbf{k}_{1},\mathbf{k}_{23})\gamma_{2}(\mathbf{k}_{2},\mathbf{k}_{3}) \label{eq: high_Poisson_2}
\end{align}
where
\begin{align}
\gamma_{2}(\mathbf{k}_{1},\mathbf{k}_{2}) =1-\frac{(\mathbf{k}_{1}\cdot\mathbf{k}_{2})^{2}}{k_{1}^{2}k_{2}^{2}}
\end{align}
and where $S_2,S_{22}$ are functions of time that depend on the specific Horndeski model. An additional cubic-order term  has been neglected because it vanishes when the gravitational wave speed equals the speed of light.

 The non-linear Poisson corrections would modify the kernels by additional terms. We will write  them down in due course for completeness, but they will not be included in the calculations.  Therefore, our results are only valid at scales larger than the Vainshtein radius.  

We now proceed to derive the second-order kernels $F_2$ and $G_2$.

\subsection{Second-Order Kernels} \label{sec:2nd_FG} %

The continuity and Euler equations at second order, after symmetrization, are given by %
\begin{align} \label{eq: 2nd_continuity}
\delta^{(2)\prime}_{\mathbf{k}} -  \theta^{(2)}_{\mathbf{k}} 
&= \frac{1}{2} \int_{\mathbf{k}_{12} = \mathbf{k}} \Big[ \alpha_{1,2}\, f_1 + \alpha_{2,1}\, f_2 \Big] \delta^{(1)}_{\mathbf{k}_1}\, \delta^{(1)}_{\mathbf{k}_2} \,,
\\
\label{eq:2nd_Euler}
\theta^{(2)\prime}_{\mathbf{k}} + \mathcal{F}\, \theta^{(2)}_{\mathbf{k}} - S(k)\, \delta^{(2)}_{\mathbf{k}} 
&= \int_{\mathbf{k}_{12} = \mathbf{k}} \beta_{1,2}\, f_1 f_2\, \delta^{(1)}_{\mathbf{k}_1}\, \delta^{(1)}_{\mathbf{k}_2} \,,
\end{align} %
where we introduce the shorthand notation \( f_i \equiv f(k_i) \). 
Including the Vainshtein corrections \cite{Cusin_2018}, would result in a new kernel
\begin{align}
\beta_{V1,2}(\mathbf{k}_{1},\mathbf{k}_{2}) & =\beta_{1,2}(\mathbf{k}_{1},\mathbf{k}_{2})+V_{2}\gamma_2(\mathbf{k}_{1},\mathbf{k}_{2})\label{eq:beta-Vain}
\end{align}
where $V_2$ is a function that depends on $S_2$ and $f_1,f_2$ \footnote{By comparing with Eq. \eqref{eq: high_Poisson_1}, one can find that $V_2(N, \mathbf{k}_1, \mathbf{k}_2)=\frac{S_2(N)}{f_1f_2}$. }. As already mentioned, we put $V_2=0$ in the following.

Substituting the ansatz from Eq.~\eqref{eq: kernels} yields a coupled system for the second-order kernels \( F_2 \) and \( G_2 \):
\begin{align} \label{eq:2nd_F_2}
F_2' + F_2(f_1 + f_2) - G_2 & = \frac{1}{2} \left( \alpha_{1,2} f_1 + \alpha_{2,1} f_2 \right) \,,
\\
\label{eq:2nd_G_2}
G_2' + G_2(f_1 + f_2) + \mathcal{F}(N) G_2 - S(k) F_2 &= \beta_{1,2} f_1 f_2 \,.
\end{align} %
 Combining Eqs.~\eqref{eq:2nd_F_2} and \eqref{eq:2nd_G_2}, and using \( f_i'=S(k_i)-\mathcal{F}f_i-f_i^2 \) , we obtain a second-order differential equation for \( F_2 \): %
\begin{align}
& F_2'' + 2\left(f_1 + f_2 + \frac{\mathcal{F}}{2}\right) F_2' + \Big[2 f_1 f_2 + S(k_1) + S(k_2) - S(k)\Big] F_2  \nonumber 
\\&= \frac{1}{2} \Big[ \alpha_{1,2} S(k_1) + \alpha_{2,1} S(k_2) \Big] 
 + \frac{1}{2} f_1 f_2 (\alpha_{1,2} + \alpha_{2,1}) + \beta_{1,2} f_1 f_2 \,.\label{eq:F_2_second_order}
\end{align} 
Following Ref.~\cite{Aviles:2021que}, we define the second-order growth function: %
\begin{equation} \label{eq:2nd_D}
D^{(2)}(\mathbf{k}_1, \mathbf{k}_2, N) \equiv D_{12} \equiv 2 D_1 D_2 F_2 - \chi_{1,2} \quad \Rightarrow \quad F_2 = \frac{D_{12}}{2 D_1 D_2} + \frac{1}{2} \chi_{1,2} \,,
\end{equation} %
with 
\begin{equation} \label{eq:chiD}
\chi_{1,2} \equiv \alpha_{1,2} + \alpha_{2,1} - \gamma_{1,2}, \qquad 
\gamma_{1,2} = 1 - \frac{(\mathbf{k}_1 \cdot \mathbf{k}_2)^2}{k_1^2 k_2^2}, \qquad 
D_i \equiv D(\mathbf{k}_i, N) \,.
\end{equation} %
This choice simplifies the structure of the second-order equations.

Substituting Eq.~\eqref{eq:2nd_D} into Eq.~\eqref{eq:F_2_second_order} allows us to recast the equation in terms of \( D_{12} \): %
\begin{align}
D_{12}'' + \mathcal{F}\, D_{12}' - S(k)\, D_{12} 
=\ &\Bigg[ S(k) + \left( S(k) - S(k_2) \right) \frac{\mathbf{k}_1 \cdot \mathbf{k}_2}{k_1^2} 
+ \left( S(k) - S(k_1) \right) \frac{\mathbf{k}_1 \cdot \mathbf{k}_2}{k_2^2} \nonumber \\
&\quad - \left( S(k_1) + S(k_2) - S(k) \right) \frac{(\mathbf{k}_1 \cdot \mathbf{k}_2)^2}{k_1^2 k_2^2} 
\Bigg] D_1 D_2 \,. \label{eq:dd_D12}
\end{align} %
The solution can be written as 
\begin{equation}
D_{12} = D_{12,\mathcal{A}} - \frac{(\mathbf{k}_1 \cdot \mathbf{k}_2)^2}{k_1^2 k_2^2}\, D_{12,\mathcal{B}}\,,
\end{equation} %
where \( D_{12,\mathcal{A}} \) and \( D_{12,\mathcal{B}} \) satisfy
\begin{align}
D_{12,\mathcal{A}}'' + \mathcal{F}\, D_{12,\mathcal{A}}' - S(k)\, D_{12,\mathcal{A}} 
&= \Bigg[ S(k) + \left( S(k) - S(k_1) \right) \frac{\mathbf{k}_1 \cdot \mathbf{k}_2}{k_2^2} 
+ \left( S(k) - S(k_2) \right) \frac{\mathbf{k}_1 \cdot \mathbf{k}_2}{k_1^2} \Bigg] D_1 D_2 \equiv \mathcal{I}_{\mathcal{A}}\,, \label{eq:D2A_N} \\
D_{12,\mathcal{B}}'' + \mathcal{F}\, D_{12,\mathcal{B}}' - S(k)\, D_{12,\mathcal{B}} 
&= \left[ S(k_1) + S(k_2) - S(k) \right] D_1 D_2 \equiv \mathcal{I}_{\mathcal{B}}\,. \label{eq:D2B_N}
\end{align}

Finally, the kernels \( F_2 \) and \( G_2 \) are obtained from Eqs.~\eqref{eq:2nd_D} and (\ref{eq:2nd_F_2},\ref{eq:2nd_G_2}): %
\begin{align}
F_2(\mathbf{k}_1, \mathbf{k}_2) 
&= \frac{1}{2} + \frac{3}{14} \mathcal{A} 
+ \left( \frac{1}{2} - \frac{3}{14} \mathcal{B} \right) \frac{(\mathbf{k}_1 \cdot \mathbf{k}_2)^2}{k_1^2 k_2^2} 
+ \frac{\mathbf{k}_1 \cdot \mathbf{k}_2}{2 k_1 k_2} \left( \frac{k_2}{k_1} + \frac{k_1}{k_2} \right) \,,
\label{eq:F2kernel} \\
G_2(\mathbf{k}_1, \mathbf{k}_2) 
&= \frac{3\mathcal{A}(f_1 + f_2) + 3\mathcal{A}'}{14}
+ \left( \frac{f_1 + f_2}{2} - \frac{3\mathcal{B}(f_1 + f_2) + 3\mathcal{B}'}{14} \right) \frac{(\mathbf{k}_1 \cdot \mathbf{k}_2)^2}{k_1^2 k_2^2}
+ \frac{\mathbf{k}_1 \cdot \mathbf{k}_2}{2 k_1 k_2} \left( \frac{f_2 k_2}{k_1} + \frac{f_1 k_1}{k_2} \right) \,,
\label{eq:F_2_G_2_kernel}
\end{align} %
with
\begin{equation} \label{eq:mathAB-1}
\mathcal{A}(\mathbf{k}_1, \mathbf{k}_2, N) = \frac{7 D_{12,\mathcal{A}}(\mathbf{k}_1, \mathbf{k}_2, N)}{3 D_1D_2} \,, \qquad 
\mathcal{B}(\mathbf{k}_1, \mathbf{k}_2, N) = \frac{7 D_{12,\mathcal{B}}(\mathbf{k}_1, \mathbf{k}_2, N)}{3 D_1D_2} \,.
\end{equation} %

\subsection{Third order kernels}\label{sec:3rd_FG} %

In this section, we derive the third-order SPT kernels directly from the fluid equations. At third order, the continuity and Euler equations take the form %
\begin{align}
\delta^{(3)\prime}_{\mathbf{k}} - \theta^{(3)}_{\mathbf{k}} &= \int_{\mathbf{k}_{12}=\mathbf{k}} \alpha_{1,2} \left( \theta^{(1)}
_{\mathbf{k}_{1}}\delta^{(2)}_{\mathbf{k}_{2}} + \theta^{(2)}_{\mathbf{k}_{1}}\delta^{(1)}_{\mathbf{k}_{2}} \right)\,, \label{eq:3rd_continuity}
\\
\theta^{(3)\prime}_{\mathbf{k}} + \mathcal{F}\, \theta^{(3)}_{\mathbf{k}} - S(k)\delta^{(3)}_{\mathbf{k}} &= \int_{\mathbf{k}_{12}=\mathbf{k}} \beta_{1,2} \left( \theta^{(1)}_{\mathbf{k}_{1}}\theta^{(2)}_{\mathbf{k}_{2}} + \theta^{(2)}_{\mathbf{k}_{1}}\theta^{(1)}_{\mathbf{k}_{2}} \right)\,, \label{eq:3rd_Euler}
\end{align}
where \( \delta^{(3)} \) and \( \theta^{(3)} \) are the third-order density contrast and velocity divergence, respectively. %

Inserting the first- and second-order kernels as defined in Eq.~\eqref{eq: kernels}, 
into the right-hand sides of Eqs.~\eqref{eq:3rd_continuity},  %
the continuity equation becomes (already symmetrized) 
\begin{align} \label{eq:full_continuity}
\delta_{\textbf{k}}^{(3)}{}' - \theta^{(3)}_{\textbf{k}} 
&= \frac{1}{3} \left\{ \int_{\textbf{k}_1+\textbf{k}_{23}=\textbf{k}} \alpha_{1,23} f_1 \delta_{\textbf{k}_1} \delta_{\textbf{k}_2}\delta_{\textbf{k}_3} F_{2}(\textbf{k}_2, \textbf{k}_3) \right\}_{\text{cyc}} \nonumber \\
&\quad + \frac{1}{3} \left\{\int_{\textbf{k}_{13} + \textbf{k}_2 = \textbf{k}} \alpha_{13,2} \delta_{\textbf{k}_1} \delta_{\textbf{k}_2} \delta_{\textbf{k}_3} G_{2}(\textbf{k}_1, \textbf{k}_3) \right\}_{\text{cyc}} \nonumber \\
&= \frac{1}{3} \int_{\textbf{k}_{123} = \textbf{k}} \hat{\alpha}(\textbf{k}_{1}, \textbf{k}_{2}, \textbf{k}_{3}) \delta_{\textbf{k}_{1}} \delta_{\textbf{k}_{2}} \delta_{\textbf{k}_{3}}  \,, 
\end{align}
where 
\begin{equation} \label{eq: hat_a}
\hat{\alpha}(\mathbf{k}_1, \mathbf{k}_2, \mathbf{k}_3)=\Big\{\alpha_{1,23}f_1F_{2}(\mathbf{k}_{2},\mathbf{k}_{3})+\alpha_{13,2}G_{2}(\mathbf{k}_{1},\mathbf{k}_{3})\Big\}_{\text{cyc}}\, ,
\end{equation} %
and \( \alpha_{1,23}=\alpha(\textbf{k}_1,\textbf{k}_{23})\) and similar notation. Here and below, we use \(\{\}_{\text{cyc}}\) to denote the sum over the three cyclic permutations of the triplet \((\mathbf{k}_1, \mathbf{k}_2, \mathbf{k}_3)\).

Likewise, from Eq.~\eqref{eq:3rd_Euler} the third-order symmetrized Euler equation is given by %
\begin{align} \label{eq:full_Euler}
\text{l.h.s.} 
&= \frac{1}{3} \left\{ \int_{\mathbf{k}_1 + \mathbf{k}_{23} = \mathbf{k}} 
\beta_{1,23} f_1 
\, G_2(\mathbf{k}_2, \mathbf{k}_3)\, \delta_{\mathbf{k}_1}\, \delta_{\mathbf{k}_2}\, \delta_{\mathbf{k}_3} \right\}_{\text{cyc}} \nonumber \\
&\quad + \frac{1}{3} \left\{ \int_{\mathbf{k}_2 + \mathbf{k}_{13} = \mathbf{k}} 
\beta_{13,2} f_2 
\, G_2(\mathbf{k}_1, \mathbf{k}_3)\, \delta_{\mathbf{k}_1}\, \delta_{\mathbf{k}_2}\, \delta_{\mathbf{k}_3} \right\}_{\text{cyc}}  \nonumber \\
&= \frac{2}{3} \int_{\mathbf{k}_{123} = \mathbf{k}} 
\hat{\beta}(\mathbf{k}_1, \mathbf{k}_2, \mathbf{k}_3)\, 
\delta_{\mathbf{k}_1}\, \delta_{\mathbf{k}_2}\, \delta_{\mathbf{k}_3} \,, 
\end{align}
where %
\begin{equation}
\hat{\beta}(\mathbf{k}_1, \mathbf{k}_2, \mathbf{k}_3) =\Big\{\beta_{1,23}f_1G_{2}(\mathbf{k}_{2},\mathbf{k}_{3}) \Big\}_{\text{cyc}} \,.\label{eq:hat_beta}
\end{equation}
For a more detailed derivation of Eqs.~\eqref{eq:full_continuity}, see App. ~\ref{sec:detailed}. 

Just as for the second order case, including the Vainshtein corrections \cite{Cusin_2018}, would result in a new kernel
\begin{align}
\beta_V(\mathbf{k}_1, \mathbf{k}_2, \mathbf{k}_3) & =\hat{\beta}(\mathbf{k}_1, \mathbf{k}_2, \mathbf{k}_3)+\frac{1}{2}V_{22}\left\{\gamma_{2}(\mathbf{k}_{1},\mathbf{k}_{23})\gamma_{2}(\mathbf{k}_{2},\mathbf{k}_{3})\right\}_{\text{cyc}}\label{eq:beta-Vain2}
\end{align}
where $V_{22}$ is a function of time that depends on $S_{22}$ \footnote{Likewise, from Eq.~(3.7), one can obtain $V_{22} (N)= S_{22}(N)$.}. We put $V_{22}=0$ in the following.

From Eq.~\eqref{eq: kernels}, the third-order density and velocity divergence fields are defined, respectively, as %
\begin{align}
\delta^{(3)}(\textbf{k}) &\equiv \int_{\textbf{k}_{123} = \textbf{k}} F_{3}(\textbf{k}_{1}, \textbf{k}_{2}, \textbf{k}_{3})\, D_{1} D_{2} D_{3}\, \delta_{0}(\textbf{k}_{1})\, \delta_{0}(\textbf{k}_{2})\, \delta_{0}(\textbf{k}_{3})\,, \label{eq:delta_3}
\\
\theta^{(3)}(\textbf{k}) &\equiv \int_{\textbf{k}_{123} = \textbf{k}} G_{3}(\textbf{k}_{1}, \textbf{k}_{2}, \textbf{k}_{3})\, D_{1} D_{2} D_{3}\, \delta_{0}(\textbf{k}_{1})\, \delta_{0}(\textbf{k}_{2})\, \delta_{0}(\textbf{k}_{3})\,, \label{eq:theta_3}
\end{align}
in which $\delta_{0}(\textbf{k}_i)=\delta^{(1)}(\textbf{k}_i,N_{0})$, and $D_i$ is defined in Eq.~\eqref{eq:chiD}

Inserting Eq.~\eqref{eq:delta_3} and \eqref{eq:theta_3} into the third %
order fluid equations Eq.~\eqref{eq:3rd_continuity} and Eq.~\eqref{eq:3rd_Euler}, %
we obtain %
\begin{align}
(F_{3}D_{1}D_{2}D_{3})'-G_{3}D_1D_2D_3&=\frac{1}{3}\hat{\alpha}D_1D_2D_3  \label{eq:dF3_dt}\,,\\ 
(G_{3}D_{1}D_{2}D_{3})'+\mathcal{F}G_{3}D_1D_2D_3-S(k)F_{3}D_1D_2D_3&=\frac{2}{3}\hat{\beta}D_1D_2D_3 \label{eq:dG3_dt} \,, %
\end{align}
where \( k=|\mathbf{k}_{1}+\mathbf{k}_{2}+\mathbf{k}_{3}| \). Using again \( f_i = D_i'/D_i \), and combining the two equations above \footnote{To derive a single equation for \( F_3 \), we take the time derivative of Eq.~\eqref{eq:dF3_dt} and substitute the expression for \( (G_3 D_1 D_2 D_3)' \) from Eq.~\eqref{eq:dG3_dt}. Moreover, we use Eq.~\eqref{eq:dF3_dt} to eliminate \( G_3 D_1 D_2 D_3 \).}, we obtain %
\begin{align}
(F_{3}D_{1}D_{2}D_{3})'' + \mathcal{F}(F_3D_{1}D_{2}D_{3})'-S(k)(F_{3}D_{1}D_{2}D_{3})= \frac{1}{3}\left[2\hat{\beta}+(f_1+f_2+f_3+\mathcal{F})\hat{\alpha}+\hat{\alpha}'\right]D_{1}D_{2}D_{3} \label{eq:F_3_dt2}\,,
\end{align}
where \( \hat{\alpha}' \) can be straightforwardly obtained from Eq.~\eqref{eq: hat_a} as %
\begin{align}
\hat{\alpha}'(\mathbf{k}_1, \mathbf{k}_2, \mathbf{k}_3) = 
\left\{\alpha_{1,23} \Big[ f'_1\, F_2(\mathbf{k}_2, \mathbf{k}_3) + f_1\, F_2'(\mathbf{k}_2, \mathbf{k}_3) \Big]
 + \alpha_{23,1}\, G_2'(\mathbf{k}_2, \mathbf{k}_3)
\right\}_{\text{cyc}} \,, \label{eq: da'dt}
\end{align} %
in which \( F_2' \) and \( G_2' \) can be obtained from Eqs.~\eqref{eq:2nd_F_2} and \eqref{eq:2nd_G_2}, and both depend on \( F_2 \) and \( G_2 \) themselves.

The third-order growth function can be defined as
\begin{equation}
D^{(3)}(\mathbf{k}_{1}, \mathbf{k}_{2}, \mathbf{k}_{3}, t) \equiv D_{123} \equiv 6 D_{1} D_{2} D_{3} F_{3} \,. \label{eq:3rd_D}
\end{equation}
Inserting Eq.~\eqref{eq:3rd_D} into Eq.~\eqref{eq:F_3_dt2}, we obtain the evolution equation for \( D_{123} \),
\begin{equation}
D_{123}'' + \mathcal{F}\,D_{123}' - S(k) \,D_{123} = 6D_1D_2D_3R; \quad R\equiv\frac{1}{3}\hat{\alpha}'+\frac{2}{3}\hat{\beta}+\frac{1}{3}\hat{\alpha}(f_{1}+f_{2}+f_{3}+\mathcal{F}) \,. \label{eq:dd_D123}
\end{equation} %
By defining %
\begin{equation}
\mathcal{A}_{3}(\mathbf{k}_{1}, \mathbf{k}_{2}, \mathbf{k}_{3}, N) = \frac{7D_{123}(\mathbf{k}_{1}, \mathbf{k}_{2}, \mathbf{k}_{3}, N)}{3D_1 D_2 D_3}\,, \qquad
\end{equation}
we obtain the third-order density and velocity kernels \( F_3 \),  \( G_3 \) from Eqs.~\eqref{eq:3rd_D} and ~\eqref{eq:dF3_dt}, %
\begin{align}
F_3 &= \frac{1}{14}\mathcal{A}_3 \,, \label{eq:final_F3}  \\
G_3 &= \frac{1}{14}\mathcal{A}_3' + \frac{1}{14}\mathcal{A}_3 
\left(f_1 + f_2 + f_3 \right) - \frac{1}{3} \hat{\alpha}\,. \label{eq:final_G3}
\end{align}
As a side note, the kernels \( F_3 \) and \( G_3 \) have been obtained via third-order Lagrangian perturbation theory in Ref.~\cite{Aviles:2021que}.

The expressions $\mathcal{A},\mathcal{B},\mathcal{A}_3$ fully determine the kernels. We now proceed to derive them in the case of scale-dependent Horndeski gravity.


\section{Kernels in Horndeski gravity}\label{sec:F2G2F3G3}

We emphasize that all results derived in Sec.~\ref{sec: kernels} apply to general models with scale-dependent growth, such as those involving massive neutrinos or modified gravity, and remain valid regardless of the specific form of the scale dependence. In this section, we specialize to Horndeski gravity, where the modifications can be captured by two functions: \( h_1 \), which depends only on time, and \( h_c \), which is both time- and scale-dependent. %

\subsection{Second-order kernels} \label{sec:free_F2G2}

We now proceed to solve Eqs.~\eqref{eq:D2A_N} and \eqref{eq:D2B_N}, which play a central role in the analysis presented in this subsection. As shown in Sec.~\ref{sec:Linear}, the quantities $S$, $D_{1}$,
and $D_{2}$ are fully determined once the background cosmology and
the Horndeski parameters $h_{1}$ and $h_{c}$ are specified. Recall from Eq.~\eqref{eq: A_k_honedeski} that the source term $S$ can be decomposed into a time-dependent component $S_{z}(N)$ and a scale-dependent
component $S_{kz}(k,N)$. This decomposition leads to the following expressions for $\mathcal{I}_{\mathcal{A}}$ and $\mathcal{I}_{\mathcal{B}}$:
\begin{align}
\mathcal{I}_{\mathcal{A}} & =\frac{3}{2}\Omega_{m}\Biggl[h_{1}+h_{c}(k)+(h_{c}(k)-h_{c}(k_{1}))\frac{\mathbf{k}_{1}\cdot\mathbf{k}_{2}}{k_{2}^{2}}+(h_{c}(k)-h_{c}(k_{2}))\frac{\mathbf{k}_{1}\cdot\mathbf{k}_{2}}{k_{1}^{2}}\Biggr]D_{1}D_{2}\,,\\
\mathcal{I}_{\mathcal{B}} & =\frac{3}{2}\Omega_{m}\Biggl[h_{1}+h_{c}(k_{1})+h_{c}(k_{2})-h_{c}(k)\Biggr]D_{1}D_{2}\,.
\end{align}

As before, we decompose \( D_{12,\mathcal{A}}(\mathbf{k}_1, \mathbf{k}_2, N) \) and \( D_{12,\mathcal{B}}(\mathbf{k}_1, \mathbf{k}_2, N) \) into two components: a purely time-dependent part, denoted \( D_{12,\mathcal{A}_z}(N) \) and \( D_{12,\mathcal{B}_z}(N) \), and a sub-dominant term with explicit scale dependence, denoted \( D_{12,\mathcal{A}_{kz}}(\mathbf{k}_1, \mathbf{k}_2, N) \) and \( D_{12,\mathcal{B}_{kz}}(\mathbf{k}_1, \mathbf{k}_2, N) \). It is straightforward to verify that \[ D_{12,\mathcal{A},z}(N) = D_{12,\mathcal{B},z}(N) \equiv D_{12,z}(N), \]
which satisfies the equation
\begin{equation}
D_{12,z}''+\mathcal{F}D_{12,z}'-\frac{3}{2}\Omega_{m}h_{1}D_{12,z}=\frac{3}{2}\Omega_{m}h_{1}D_{z}^{2}\equiv\mathcal{I}_{z}\,,\label{eq:d12z}
\end{equation}
where the source term $\mathcal{I}_{z}$ depends quadratically on
the linear growth function $D_{z}$. Note that the source term $\mathcal{I}_{z}$ is proportional to $D_{z}^2$, whereas the source in Eq.~\eqref{eq:dkz-1} for $D_{kz}$ depends linearly on $D_{z}$.

Equation~\eqref{eq:d12z} can be solved using the Wronskian method.
Let us denote the two linearly independent solutions to the associated
homogeneous equation as $D_{+}(N)\,, D_{-}(N)\,$, which coincide with the linear modes,  since they obey the same equation. A particular solution to the inhomogeneous equation is then given by
\begin{align}
D_{12,z}(N) &= -D_{+}(N)\int_{N_{0}}^{N}\mathrm{d}x\,\frac{D_{-}(x)\,\mathcal{I}_{z}(x)}{W(D_{+},D_{-})} + D_{-}(N)\int_{N_{0}}^{N}\mathrm{d}x\,\frac{D_{+}(x)\,\mathcal{I}_{z}(x)}{W(D_{+},D_{-})} \nonumber\\
&= \frac{3}{2}\left[-D_{+}(N)\int_{N_{0}}^{N}\mathrm{d}x\,\frac{\Omega_{m}(x) D_{-}(x) D_{+}^{2}(x) h_{1}(x)}{W(D_{+},D_{-})} + D_{-}(N)\int_{N_{0}}^{N}\mathrm{d}x\,\frac{\Omega_{m}(x) D_{+}^{3}(x) h_{1}(x)}{W(D_{+},D_{-})}\right]\,.
\label{eq:D_12_z_Wronskian}
\end{align}
    
Keeping terms up to the first order in $h_{c}$,
we can further derive two equations for $D_{12,\mathcal{A}_{kz}}$
and $D_{12,\mathcal{B}_{kz}}$: 
\begin{align}
D_{12,\mathcal{A}_{kz}}''+\mathcal{F}D_{12,\mathcal{A}_{kz}}'-\frac{3}{2}\Omega_{m}h_{1}D_{12,\mathcal{A}_{kz}} & =\mathcal{I}_{\mathcal{A}}-\mathcal{I}_{z}+S_{kz}D_{12,z}\equiv\hat{\mathcal{I}}_{\mathcal{A}}\,,\label{eq:DeltaA2_N}\\
D_{12,\mathcal{B}_{kz}}''+\mathcal{F}D_{12,\mathcal{B}_{kz}}'-\frac{3}{2}\Omega_{m}h_{1}D_{12,\mathcal{B}_{kz}} & =\mathcal{I}_{\mathcal{B}}-\mathcal{I}_{z}+S_{kz}D_{12,z}\equiv\hat{\mathcal{I}}_{\mathcal{B}}\,.\label{eq:DeltaB2_N}
\end{align} %
Using the decomposition \( D_i = D_z + D_{kz}(k_i) \) (here $D_z$ can be identified with the linear growing mode $D_+$), where we suppress the explicit time dependence, the source terms \( \hat{\mathcal{I}}_{\mathcal{A}} \) and \( \hat{\mathcal{I}}_{\mathcal{B}} \) are given by

\begin{align}
\hat{\mathcal{I}}_{\mathcal{A}} & \equiv\frac{3}{2}\Omega_{m}\left\{ h_{1}D_z(D_{kz}(k_{1})+D_{kz}(k_{2}))\right.\\&\left.+\left[h_{c}(k)+(h_{c}(k)-h_{c}(k_{1}))\frac{\mathbf{k}_{1}\cdot\mathbf{k}_{2}}{k_{2}^{2}}+(h_{c}(k)-h_{c}(k_{2}))\frac{\mathbf{k}_{1}\cdot\mathbf{k}_{2}}{k_{1}^{2}}\right]D_{z}^{2}+h_{c}(k)D_{12,z}\right\} \nonumber\\
\hat{\mathcal{I}}_{\mathcal{B}} & \equiv\frac{3}{2}\Omega_{m}\left\{ h_{1}D_z(D_{kz}(k_{1})+D_{kz}(k_{2}))+\Biggl[h_{c}(k_{1})+h_{c}(k_{2})-h_{c}(k)\Biggr]D_{z}^{2}+h_{c}(k)D_{12,z}\right\}. \label{eq:iab-1}
\end{align}

Eq.~\eqref{eq:DeltaA2_N} and Eq.\eqref{eq:DeltaB2_N} can also be
solved using the Wronskian method. The particular solution to the
inhomogeneous equation is given by 
\begin{equation}
D_{12,\mathcal{A}_{kz}}(\textbf{k}_{1},\textbf{k}_{2},N)=-D_{+}(N)\int_{N_{0}}^{N}\mathrm{d}x\frac{D_{-}(x)\hat{\mathcal{I}}_{\mathcal{A}}(x)}{W\left(D_{+}(x),D_{-}(x)\right)}+D_{-}(N)\int_{N_{0}}^{N}\mathrm{d}x\frac{D_{+}(x)\hat{\mathcal{I}}_{\mathcal{A}}(x)}{W\left(D_{+}(x),D_{-}(x)\right)}\,,\label{eq:par_solu-a}
\end{equation}
and similarly for \( D_{12,\mathcal{B}_{kz}} \).

By construction, $D_{12,\mathcal{A/B}}=D_{12,z}+D_{12,\mathcal{A}_{kz}/\mathcal{B}_{kz}}$.
Once $D_{12,\mathcal{A/B}}$ is obtained, one can derive $F_{2}$
and $G_{2}$ by substituting $\mathcal{A}$ and $\mathcal{B}$ as
defined in Eq.~\eqref{eq:mathAB-1}. To first order in $h_{c}$,  $\mathcal{A}$ and $\mathcal{B}$ are given as follows 
\begin{align}
\mathcal{A} &  =\frac{7D_{12,z}}{3D_{+}^{2}}\left[1+\int_{N_{0}}^{N}\mathrm{d}x\frac{D_{-}D_{+} \big(S_{kz}(k_1)+S_{kz}(k_2)\big)}{W} - \frac{D_{-}}{D_{+}}\int_{N_{0}}^{N}\mathrm{d}x\frac{D_{+}^{2} \big(S_{kz}(k_1)+S_{kz}(k_2)\big)}{W}\right]+\nonumber\\
 & \frac{7}{3}\left\{ -\frac{1}{D_{+}}\int_{N_{0}}^{N}\mathrm{d}x\frac{D_{-}\hat{\mathcal{I}}_{\mathcal{A}}}{W}+\frac{D_{-}}{D_{+}^{2}}\int_{N_{0}}^{N}\mathrm{d}x\frac{D_{+}\hat{\mathcal{I}}_{\mathcal{A}}}{W}\right\} \label{eq:Horndeski_a}  \,,
\end{align}
and 
\begin{align}
\mathcal{B} & =\frac{7D_{12,z}}{3D_{+}^{2}}\left[1+\int_{N_{0}}^{N}\mathrm{d}x\frac{D_{-}D_{+} \big(S_{kz}(k_1)+S_{kz}(k_2)\big)}{W} - \frac{D_{-}}{D_{+}}\int_{N_{0}}^{N}\mathrm{d}x\frac{D_{+}^{2} \big(S_{kz}(k_1)+S_{kz}(k_2)\big)}{W}\right]+ \nonumber\\
 &  \frac{7}{3}\left\{ -\frac{1}{D_{+}}\int_{N_{0}}^{N}\mathrm{d}x\frac{D_{-}\hat{\mathcal{I}}_{\mathcal{B}}}{W}+\frac{D_{-}}{D_{+}^{2}}\int_{N_{0}}^{N}\mathrm{d}x\frac{D_{+}\hat{\mathcal{I}}_{\mathcal{B}}}{W}\right\} \label{eq:Horndeski_b}
 \,.
\end{align} %
We see therefore that the Horndeski kernels at first order in $h_{c}$
are entirely determined in terms of the $k$-independent  linear
growth function $D_{+}$.

\subsection{Third-order kernels} \label{sec:free_F3G3}  

We now proceed to solve Eqs.~\eqref{eq:dd_D123}, which is central to constructing the third-order kernels. As its structure is analogous to that of Eqs.~\eqref{eq:D2A_N} and \eqref{eq:D2B_N}, the same solution method applies. Thus, it remains only to derive explicit expressions for \( D_{123} \).

As in the second-order case discussed in Sec.~\ref{sec:free_F2G2}, we decompose \( D_{123}(k,N) \) into a leading term \( D_{123,z} \), arising from the purely time-dependent growth, and a subleading, scale-dependent correction \( D_{123,kz} \). We compute both contributions accordingly and, for completeness, provide the differential equation governing \( D_{123,z} \).

The evolution equation for $D_{123,z}$ reads: 
\begin{align}
D_{123,z}''+\mathcal{F}D_{123,z}'-\frac{3}{2}\Omega_{m}h_{1}D_{123,z}&=R_{h_c^0}6D_{z}^{3}\equiv\mathcal{I}_{3,z}\,,\label{eq:LCDM_DA^3_1}
\end{align}
where \( R_{h_c^0} \) denotes the component of \( R \) at order \( h_c^0 \), given by 
\begin{equation} \label{eq:R_1z}
    R_{h_c^0}=   \frac{1}{3}\hat{\alpha}_{h_c^0}'+ \frac{2}{3}\hat{\beta}_{h_c^0}+\frac{1}{3}\hat{\alpha}_{h_c^0}(3f_z+\mathcal{F})   \,.
\end{equation} %
A more explicit expression of $R_{h_c^0}$ is provided in App. ~\ref{sec:detailed}. The solution for $D_{123,z}$ is:
\begin{equation}
D_{123,z}(N)=-D_{+}(N)\int_{N_{0}}^{N}\mathrm{d}x\frac{D_{-}(x)\mathcal{I}_{3,z}(x)}{W\left(D_{+}(x),D_{-}(x)\right)}+D_{-}(N)\int_{N_{0}}^{N}\mathrm{d}x\frac{D_{+}(x)\mathcal{I}_{3,z}(x)}{W\left(D_{+}(x),D_{-}(x)\right)}\,.\label{eq:par_solu_third_z}
\end{equation}

Similarly, to first order in $h_c$, the equation for $D_{123,kz}$ satisfies
\begin{equation}
D_{123,kz}''+\mathcal{F}D_{123,kz}'-\frac{3}{2}\Omega_{m}h_{1}D_{123,kz}\equiv\hat{\mathcal{I}}_{3}\,,\label{eq:LCDM_DA^3}
\end{equation}
with source term 
\begin{equation}
\hat{\mathcal{I}}_{3}=6D_1D_2D_3R-\mathcal{I}_{3,z}+S_{kz}D_{123,z}\,,
\end{equation}
and corresponding solution
\begin{equation}
D_{123,kz}(\textbf{k}_{1},\textbf{k}_{2},\textbf{k}_{3},N)=-D_{+}(N)\int_{N_{0}}^{N}\mathrm{d}x\frac{D_{-}(x)\hat{\mathcal{I}}_{3}(x)}{W\left(D_{+}(x),D_{-}(x)\right)}+D_{-}(N)\int_{N_{0}}^{N}\mathrm{d}x\frac{D_{+}(x)\hat{\mathcal{I}}_{3}(x)}{W\left(D_{+}(x),D_{-}(x)\right)}\,. \label{eq:par_solu_third}
\end{equation} %
Following the procedure in Sec.~\ref{sec:free_F2G2}, the source $\hat{\mathcal{I}}_3$ to first order in $h_{c}$ is given by
\begin{align}
\hat{\mathcal{I}}_3
&=\Big(2\hat{\alpha}'_{h_c^0}+4\hat{\beta}_{h_c^0}+2\hat{\alpha}_{h_c^0}(3f_z+\mathcal{F})\Big)\Big[D_{kz}(k_1)+D_{kz}(k_2)+D_{kz}(k_3)\Big]D_z^2+2\hat{\alpha}'_{h_c^1}D_z^3 +4\hat{\beta}_{h_c^1}D_z^3 \nonumber  \\
 &+2\hat{\alpha}_{h_c^1}(3f_z+\mathcal{F})D_z^3 + 2\hat{\alpha}_{h_c^0}\Big[f_{kz}(k_1)+f_{kz}(k_2)+f_{kz}(k_3)\Big]D_z^3+\frac{3}{2}\Omega_{m}h_{c}D_{123,z}  \,,  \label{eq:i3a-1}
\end{align}
where the subscripts \( h_c^0 \) and \( h_c^1 \) refer to the leading and next-to-leading contributions in the expansion of \( h_c \) in \( \hat{\alpha} \) and \( \hat{\beta} \). Explicit expressions of these quantities are provided in App. ~\ref{sec:detailed}. Then, $\mathcal{A}_3$, to first order in $h_c$, is:
\begin{align}
\mathcal{A}_3 &  =\frac{7D_{123,z}}{3D_{+}^{3}}\left[1+\int_{N_{0}}^{N}\mathrm{d}x\frac{D_{-}D_{+} \big(S_{kz}(k_1)+S_{kz}(k_2)+S_{kz}(k_3)\big)}{W} \right.\\&- \left. \frac{D_{-}}{D_{+}}\int_{N_{0}}^{N}\mathrm{d}x\frac{D_{+}^{2} \big(S_{kz}(k_1)+S_{kz}(k_2)+S_{kz}(k_3)\big)}{W}\right]\nonumber\\
 & +\frac{7}{3}\left\{ -\frac{1}{D_{+}}\int_{N_{0}}^{N}\mathrm{d}x\frac{D_{-}\hat{\mathcal{I}}_{3}}{W}+\frac{D_{-}}{D_{+}^{2}}\int_{N_{0}}^{N}\mathrm{d}x\frac{D_{+}\hat{\mathcal{I}}_{3}}{W}\right\} \label{eq:Horndeski_3a}  \,.
\end{align} 
Inserting $\mathcal{A}_3$ into Eqs. \eqref{eq:final_F3} and  \eqref{eq:final_G3}, we obtain the third-order kernels $F_3$ and $G_3$. This concludes our derivation of the Horndeski kernels. To first order in $h_c$ they are essentially analytic, up to simple one-dimensional time integrals. In App. \ref{sec:summary} we collect the main results of this section. In App. \ref{app:sep} we derive the form of the kernels in the case in which $S(k,z)$ is separable. This form helps making the numerical calculation much faster.

In the next sections we include redshift distortion and bias following the usual treatment but keeping the $k$-dependent growth, and finally assemble everything into the one-loop power spectrum.


\section{Including bias and RSD}
\label{sec: bias_RSD}

This section is based on the standard perturbation theory approach to RSD, see e.g. \cite{1999ApJ...517..531S,PhysRevD.96.043526} and on the  general bias expansion present in \cite{Aviles:2020wme}. 
To take into account the RSD effect, we need to map real space into redshift space (subscript $s$) \cite{1999ApJ...517..531S,PhysRevD.96.043526}.
\begin{align}
\delta_{s}(k)&=\int\mathrm{d}^{3}s[\frac{\delta(r)-\frac{\mathrm{d}u}{\mathrm{d}r}}{1+\frac{\mathrm{d}u}{\mathrm{d}r}}]e^{-i\mathbf{k}\mathbf{s}}=\int\frac{\mathrm{d}^{3}s}{1+\frac{\mathrm{d}u}{\mathrm{d}r}}[\delta(r)-\frac{\mathrm{d}u}{\mathrm{d}r}]e^{-i\mathbf{k}\mathbf{r}-i\mathbf{k}\frac{\mathbf{r}}{r}u}
\\&=\int\mathrm{d}^{3}r[\delta(r)-\frac{\mathrm{d}u}{\mathrm{d}r}]e^{-i\mathbf{k}\mathbf{r}-i\mathbf{k}\frac{\mathbf{r}}{r}u} \, , \label{eq:rsd}
\end{align}
where 
\begin{equation}
u=\frac{\mathbf{v}}{\mathcal{H}}\cdot\frac{\mathbf{r}}{r}
\end{equation}
is the line-of-sight velocity in units of $\mathcal{H}\equiv aH$. We write $\mathbf{v}\cdot\frac{\mathbf{r}}{r}=v\mu_{\theta}$ and employ the definition $\theta=-ik_{\theta}v/\mathcal{H}$,
so that 
\begin{equation}
e^{-i\mathbf{k}\frac{\mathbf{r}}{r}u}=e^{-i\mathbf{k}\frac{\mathbf{r}}{r}\frac{v}{\mathcal{H}}\mu_{\theta}}=e^{-ik\mu\frac{v}{\mathcal{H}}\mu_{\theta}}=e^{\frac{k}{k_{\theta}}\theta\mu\mu_{\theta}} \, .
\end{equation}
We begin by expanding Eq.~\eqref{eq:rsd} in a series of Fourier integrals,
\begin{align}
e^{k\mu\theta\frac{\mu_{\theta}}{k_{\theta}}} & =\sum_{n=0}\frac{(k\mu)^{n}}{n!}[\frac{\mu_{\theta}}{k_{\theta}}\theta(\mathbf{r})]^{n}=1+\nonumber \\
 & \sum_{n=1}\frac{(k\mu)^{n}}{n!}\int\frac{\mathrm{d}^{3}k_{1}}{(2\pi)^{3}}\frac{\mu_{1}}{k_{1}}\theta(\mathbf{k}_{1})e^{-i\mathbf{k}_{1}\mathbf{r}}\int\frac{\mathrm{d}^{3}k_{2}}{(2\pi)^{3}}\frac{\mu_{2}}{k_{2}}\theta(\mathbf{k}_{2})e^{-i\mathbf{k}_{2}\mathbf{r}}...\int\frac{\mathrm{d}^{3}k_{n}}{(2\pi)^{3}}\frac{\mu_{n}}{k_{n}}\theta(\mathbf{k}_{n})e^{-i\mathbf{k}_{n}\mathbf{r}}\\
 & =1+\sum_{n=1}\frac{(k\mu)^{n}}{n!}\int\frac{\mathrm{d}^{3}k_{1}}{(2\pi)^{3}}\frac{\mu_{1}}{k_{1}}\theta(\mathbf{k}_{1})\int\frac{\mathrm{d}^{3}k_{2}}{(2\pi)^{3}}\frac{\mu_{2}}{k_{2}}\theta(\mathbf{k}_{2})...\int\frac{\mathrm{d}^{3}k_{n}}{(2\pi)^{3}}\frac{\mu_{n}}{k_{n}}\theta(\mathbf{k}_{n})e^{-i\sum_{i}^{n}\mathbf{k}_{i}\mathbf{r}} \, ,
\end{align}
so that 
\begin{align}
\delta_{s}(\mathbf{k}) & =\int\mathrm{d}^{3}r[\delta(\mathbf{r})-\frac{\mathrm{d}u}{\mathrm{d}r}]\times \nonumber \\&
\{e^{-i\mathbf{k}\mathbf{r}}+\sum_{n=1}\frac{(k\mu)^{n}}{n!}\int\frac{\mathrm{d}^{3}k_{1}}{(2\pi)^{3}}\frac{\mu_{1}}{k_{1}}\theta(\mathbf{k}_{1})\int\frac{\mathrm{d}^{3}k_{2}}{(2\pi)^{3}}\frac{\mu_{2}}{k_{2}}\theta(\mathbf{k}_{2})..\int\frac{\mathrm{d}^{3}k_{n}}{(2\pi)^{3}}\frac{\mu_{n}}{k_{n}}\theta(\mathbf{k}_{n})e^{-i(\mathbf{k}-\sum_{i}^{n}\mathbf{k}_{i})\mathbf{r}}\} \, . \label{eq:expds}
\end{align}
In the flat-field approximation, we assume that the angle $\mu$ is constant. To incorporate galaxy-matter bias, we expand the density contrast and velocity divergence in real space. Within the framework of modified gravity cosmology, we adopt the general bias expansion established in Ref.\ \cite{Aviles:2020wme}, which has been adapted to beyond-$\Lambda \text{CDM}$ cosmologies by incorporating higher-order curvature terms.
To illustrate the procedure, we start with the terms that affect the linear theory:
\begin{equation}
\delta_{g}(\mathbf{r})=b_{1}\delta(\mathbf{r})+b_{\nabla^2\delta} \nabla^2 \delta (\mathbf{r})+...,
\label{eq: bias_expanson_first_order_delta}
\end{equation}
and
\begin{align}
    \theta_g (\mathbf{r})& = \theta(\mathbf{r})+b_{\nabla^2\theta}\nabla^2\theta(\mathbf{r}),
    \label{eq: bias_expanson_full_theta}
\end{align}
where the subscript $g$ denotes galaxies, and all bias parameters depend only on time and not on space.~\footnote{It is important to note that these high-curvature terms differ from standard higher-derivative terms, as they encapsulate the effect of the scale-dependent source in the Poisson equation and thus formally contribute at the leading order of perturbation theory.}
We now use the linear theory relation:
\begin{equation}
\mathbf{v}(\mathbf{k})=i\mathcal{H}\delta_{k}f\frac{\mathbf{k}}{k^{2}}\label{eq:vdlin-2}
\end{equation}
where we now deal with a $k$-dependent growth rate $f=f(\mathbf{k})$, and
\begin{equation}
u(r)=\frac{{\textbf{r}}}{r}\cdot\frac{{\textbf{v}(\mathbf{r})}}{\mathcal{H}}=if\int\frac{\mathrm{d}^{3}k'}{(2\pi)^{3}}\delta(\mathbf{k'})\frac{{\mathbf{k'r}}}{k'{}^{2}r}e^{i{\textbf{k}'}\cdot{\textbf{r}}}.
\end{equation}
Therefore,
\begin{align}
    \int\mathrm{d}^{3}r\frac{\mathrm{d}u}{\mathrm{d}r}e^{-i\mathbf{k}\mathbf{r}}
    &=-f\int\mathrm{d}^{3}r\frac{\mathrm{d}^{3}k}{(2\pi)^{3}}\delta(\mathbf{k}')e^{-i({\textbf{k}}-\mathbf{k}'){\textbf{r}}}\mu^{2}\\
    &=-f \int_{\mathbf{k}'=\mathbf{k}}\delta(\mathbf{k}')\mu^{2}=-f\mu^{2}\delta(\mathbf{k})=-\mu^{2}\theta(\mathbf{k}).
\end{align}
By substituting $\delta(\mathbf{r})$ in Eq.~\eqref{eq:expds} with $\delta_{g}(\mathbf{r})$ from Eq.~\eqref{eq: bias_expanson_first_order_delta}, the first terms of the expansion~\eqref{eq:expds} become
\begin{align}
\delta_{gs}(\mathbf{k})&=\int\mathrm{d}^{3}r\left[\delta_g(\mathbf{r})-\frac{\mathrm{d}u}{\mathrm{d}r}\right]e^{-i\mathbf{k}\mathbf{r}}\\
&=\int\mathrm{d}^{3}r \delta_g(\mathbf{r})e^{-i\mathbf{k}\mathbf{r}}-\int\mathrm{d}^{3}r\frac{du}{dr}e^{-i\mathbf{k}\mathbf{r}}\\
 & =\delta_g(\mathbf{k})+ \mu^2\theta_g(\mathbf{k})\label{eq:lindg},
\end{align}
where the subscript $gs$ stands for galaxies in redshift space. From Eq.~\eqref{eq: bias_expanson_first_order_delta} and ~\eqref{eq: bias_expanson_full_theta} we obtain $\delta_{gs}$ at the first order: 
\begin{align}
\delta_{gs}^{(1)}(\mathbf{k}) 
 & =\left[b_1 - b_{\nabla^2\delta} k^2 +f\mu^2(1-b_{\nabla^2\theta} k^2)\right]\delta^{(1)}(\mathbf{k}).
\label{eq:lindg_1_order}
\end{align}
The second term in Eq.~\eqref{eq:expds} gives 
\begin{align}
\delta_{gs}(\mathbf{k}) & =\int\mathrm{d}^{3}r[\delta_g(\mathbf{r})-\frac{\mathrm{d}u}{\mathrm{d}r}]e^{-i(\mathbf{k}-\mathbf{k}_{1})\mathbf{r}}k\mu\frac{\mathrm{d}^{3}k_{1}}{(2\pi)^{3}}\frac{\mu_{1}}{k_{1}}\theta_g(\mathbf{k}_{1}) \\
 & =\int\mathrm{d}^{3}r\int\frac{d^{3}k_{0}}{(2\pi)^{3}}\delta_g(\mathbf{k}_{0})e^{-i(\mathbf{k}-\mathbf{k}_{0}-\mathbf{k}_{1})\mathbf{r}}k\mu\frac{\mathrm{d}^{3}k_{1}}{(2\pi)^{3}}\frac{\mu_{1}}{k_{1}}\theta_g(\mathbf{k}_{1}) \nonumber  \\
 &\quad -\int\mathrm{d}^{3}r\frac{\mathrm{d}u}{\mathrm{d}r}e^{-i(\mathbf{k}-\mathbf{k}_{1})\mathbf{r}}k\mu\frac{\mathrm{d}^{3}k_{1}}{(2\pi)^{3}}\frac{\mu_{1}}{k_{1}}\theta_g(\mathbf{k}_{1}) \\
 & =\int\frac{d^{3}k_{0}}{(2\pi)^{3}}\frac{\mathrm{d}^{3}k_{1}}{(2\pi)^{3}}\delta_g(\mathbf{k}_{0})(2\pi)^{3}\delta_{D}(\mathbf{k}-\mathbf{k}_{0}-\mathbf{k}_{1}) k\mu\frac{\mu_{1}}{k_{1}}\theta_g(\mathbf{k}_{1}) \nonumber  \\
 &\quad +\int\mu_0^{2}\theta_g(\mathbf{k}-\mathbf{k}_{1})k\mu\frac{\mathrm{d}^{3}k_{1}}{(2\pi)^{3}}\frac{\mu_{1}}{k_{1}}\theta_g(\mathbf{k}_{1})\\
 & =\int_{\mathbf{k}_{01}=\mathbf{k}}\delta_g(\mathbf{k}_{0})k\mu\frac{\mu_{1}}{k_{1}}\theta_g(\mathbf{k}_{1})+\int_{\mathbf{k}_{01}=\mathbf{k}}\mu_0^{2}\theta_g(\mathbf{k}_{0})k\mu\frac{\mu_{1}}{k_{1}}\theta_g(\mathbf{k}_{1})\\
 & =\int_{\mathbf{k}_{01}=\mathbf{k}}[\delta_g(\mathbf{k}_{0})+\theta_g(\mathbf{k}_{0})\mu_{0}^{2}]k\mu\frac{\mu_{1}}{k_{1}}\theta_g(\mathbf{k}_{1})
\end{align}
Relabeling $0,1$,... into 1,2,..., it is then not difficult to see that the entire series can be recast in the more symmetric form 
\begin{align}
\delta_{gs}(\mathbf{k}) & =\sum_{n=1}\int_{\sum_{i=1}^{n}\mathbf{k}_{i}=\mathbf{k}}[\delta_g(\mathbf{k_{1}})+\theta_g(\mathbf{k_{1}})\mu_{1}^{2}]\frac{(k\mu)^{n-1}}{(n-1)!}\frac{\mu_{2}}{k_{2}}\theta_g(\mathbf{k}_{2})\frac{\mu_{3}}{k_{3}}\theta_g(\mathbf{k}_{3})...\frac{\mu_{n}}{k_{n}}\theta_g(\mathbf{k}_{n})\label{eq:dgs}
\end{align} 
(with the understanding that for $n=1$ the product of $\mu_{i}k_{i}^{-1}\theta_{i}$ factors reduces to unity). 

A  general form of the galaxy bias to third order can then be taken as \cite{Aviles:2020wme}: 
\begin{align}
\delta_g &= b_1 \delta +b_{\nabla^2\delta} \nabla^2 \delta + \underset{2nd}{\underbrace{ \frac{b_2}{2} \delta^2 + \frac{b_{s^2}}{2} s^2+ b_\psi \psi}} + \underset{3rd}{\underbrace{\frac{b_3}{3!} \delta^3 + \frac{b_{\delta s^2}}{2} \delta s^2  + b_{st} s t + \frac{b_{s^3}}{2} s^3}} + ...,
\label{eq: bias_expanson_full_delta}
\end{align}
where $s,t,\psi$ are defined below. We maintain $\theta_g$ as defined in Eq.~\eqref{eq: bias_expanson_full_theta} at linear level, since we consider only the linear-order contribution of the galaxy velocity divergence field.
Strictly speaking, expression~\eqref{eq: bias_expanson_full_delta} is not complete for modified gravity theories with extra degrees of freedom, since it neglects higher-curvature terms such as $\nabla^4\delta$, $\nabla^6\delta$, ... . Although these higher-curvature operators formally appear at linear order, their contributions are suppressed as long as the $k$-dependent part remains subdominant in the regime considered here, i.e., $S_{kz} \ll S_z$ (see eq.~\eqref{sources}). This condition ensures the validity of the perturbative expansion.
We construct second- and third-order scalar as $s^2 = s_{ij}s_{ij}$, $st = s_{ij}t_{ij}$ and $s^3=s_{ij}s_{jk}s_{ki}$, where the  anisotropic tensors are defined as:
\begin{align}
s_{ij}(\mathbf{k}) &= \left( \frac{k_i k_j}{k^2} - \frac{1}{3}\delta_{ij} \right) \delta(\mathbf{k})\\
t_{ij}(\mathbf{k}) &= \left( \frac{k_i k_j}{k^2} - \frac{1}{3} \delta_{ij} \right) \eta(\mathbf{k})
\end{align}
Here, $s_{ij}(\mathbf{k})$ represents  the anisotropic part of the matter density field in Fourier space and $t_{ij}(\mathbf{k})$ applies the same anisotropic projection to the velocity-density difference field $\eta(\mathbf{k})$, which is:
\begin{equation}
\eta(\mathbf{k}) = \theta(\mathbf{k}) - f(k) \delta(\mathbf{k}).
\label{eq:eta_bias}
\end{equation}
Since $s_{ij}$ appears at first order while $\eta$ vanishes at the linear level in PT, the scalar $s^2$ contributes at second order, whereas $st$ only emerges at third order. Additionally, we consider $\psi$ as a field variable that has no contribution at first order but appears starting from second order:
\begin{equation}
\psi(\mathbf{k}) \equiv \eta(\mathbf{k}) +f(k)\left(- \frac{2}{7} s^2(\mathbf{k}) + \frac{4}{21} \delta^2(\mathbf{k}) \right).
\end{equation}
By applying Eq.~\eqref{eq: bias_expanson_full_delta}, we can produce a more general kernel form compared with the standard PT in \cite{2020PhRvD.102f3533C}.
In Fourier space, the first-order fields can be found from Eq. \eqref{eq:lindg_1_order} 
\begin{align}
    \delta_{g}^{(1)}({\mathbf{k}})&=(b_{1}-b_{\nabla^{2}\delta}k^{2})\delta^{(1)}(\mathbf{k})\, ,\\
    \theta_{g}^{(1)}(\mathbf{k})&=(1-b_{\nabla^{2}\theta}k^{2})f\delta^{(1)}(\mathbf{k})  \,  .
\end{align}
According to Eq.~\eqref{eq:dgs}, the second-order terms are 
\begin{align}
\delta_{gs}^{(2)}(\mathbf{k})&= \delta_g^{(2)}(\mathbf{k})+\theta_g^{(2)}(\mathbf{k})\mu^{2}\nonumber \\
&+\int_{\mathbf{k}_{12}=\mathbf{k}}[\delta_g^{(1)}(\mathbf{k_{1}})+\theta_g^{(1)}(\mathbf{k_{1}})\mu_{1}^{2}]k\mu\frac{\mu_{2}}{k_{2}}\theta_g^{(1)}(\mathbf{k}_{2}).
\label{eq:dgs_2nd}
\end{align}
We can expand $\delta_{g}$ to  second order as:
\begin{align}
    \delta_{g}^{(2)}(\mathbf{k})
    &=\left(b_1-b_{\nabla^{2}\delta}k^{2}-b_{\psi}f\right)\delta^{(2)}(\mathbf{k})+b_{\psi}\theta^{(2)}(\mathbf{k})\nonumber \\
    &\quad+\left(\frac{1}{2}b_{2}+\frac{4}{21}b_{\psi}f\right)\int_{\mathbf{k}_{12}=\mathbf{k}}\delta^{(1)}(\mathbf{k}_{1})\delta^{(1)}(\mathbf{k}_{2})\nonumber \\
    &\quad + \left(\frac{b_{s^2}}{2}-\frac{2}{7}b_{\psi}f\right)\int_{\mathbf{k}_{12}=\mathbf{k}}\delta^{(1)}(\mathbf{k}_1)\delta^{(1)}(\mathbf{k}_2) S_{1}(\mathbf{k}_{1},\mathbf{k}_{2})\\
     \theta_{g}^{(2)}(\mathbf{k})&=(1-b_{\nabla^{2}\theta}k^{2})\theta^{(2)}(\mathbf{k})
    \label{eq:delta_theta_g_2rd_order}
\end{align}
where
\begin{equation}
S_{1}(\mathbf{k}_{1},\mathbf{k}_{2})=\frac{(\mathbf{k}_{1}\cdot\mathbf{k}_{2})^{2}}{k_{1}^{2}k_{2}^{2}}-\frac{1}{3}.
\end{equation}
Then we plug Eq.~\eqref{eq:delta_theta_g_2rd_order} into Eq.~\eqref{eq:dgs_2nd} and obtain
\begin{align}
\delta_{gs}^{(2)}(\mathbf{k}) 
& =\left(b_1-b_{\nabla^{2}\delta}k^{2}-b_{\psi}f\right)\delta^{(2)}(\mathbf{k})
+\left[(1-b_{\nabla^{2}\theta}k^{2})\mu^{2}+b_{\psi}\right]\theta^{(2)}(\mathbf{k})\nonumber \\
&+\left(\frac{1}{2}b_{2}+\frac{4}{21}b_{\psi}f\right)\int_{\mathbf{k}_{12}=\mathbf{k}}\delta^{(1)}(\mathbf{k}_{1})\delta^{(1)}(\mathbf{k}_{2})\nonumber \\
&+\left(\frac{b_{s^2}}{2}-\frac{2}{7}b_{\psi}f\right)\int_{\mathbf{k}_{12}=\mathbf{k}}S_{1}(\mathbf{k}_{1},\mathbf{k}_{2})\delta^{(1)}(\mathbf{k}_{1})\delta^{(1)}(\mathbf{k}_{2})\nonumber \\
 & +\int_{\mathbf{k}_{12}=\mathbf{k}}\left[(b_{1}-b_{\nabla^{2}\delta}k^{2})\delta^{(1)}(\mathbf{k}_{1})+(1-b_{\nabla^{2}\theta}k^{2})\mu_{1}^{2}\theta^{(1)}(\mathbf{k}_{1})\right]k\mu\frac{\mu_{2}}{k_{2}}(1-b_{\nabla^{2}\theta}k^2)\theta^{(1)}(\mathbf{k}_{2})\\
 & =\int_{\mathbf{k}_{12}=\mathbf{k}}\delta^{(1)}(\mathbf{k}_{1})\delta^{(1)}(\mathbf{k}_{2})Z_{2}(\mathbf{k}_{1},\mathbf{k}_{2}).
 \label{eq:quaddg}
\end{align}
So we have
\begin{align}
Z_{2}(\mathbf{k}_{1},\mathbf{k}_{2}) &= \left(b_1-b_{\nabla^{2}\delta}k^{2}-b_{\psi}f\right)F_{2}(\mathbf{k}_1,\mathbf{k}_2) + \left[(1-b_{\nabla^{2}\theta}k^{2})f\mu^{2}+b_{\psi}\right]G_{2}(\mathbf{k}_1,\mathbf{k}_2) \nonumber \\
&\quad + \left(\frac{b_{2}}{2}+\frac{4}{21}b_{\psi}f\right)+\left(\frac{b_{s^2}}{2}-\frac{2}{7}b_{\psi}f\right)S_{1}(\mathbf{k}_{1},\mathbf{k}_{2}) \nonumber \\
&\quad + f_{2}k\mu\frac{\mu_{2}}{k_{2}}(1-b_{\nabla^{2}\theta}k^{2}) \left[\left(b_1-b_{\nabla^{2}\delta}k^{2}\right)+f_{1}\mu_{1}^{2}(1-b_{\nabla^{2}\theta}k^{2})\right]
\label{eq: Z2_MG_unsym}
\end{align}
where $f_{i}=f(k_i)$, $f=f(k)$, and $\mu_i = \frac{\mathbf{k}_i \cdot \hat{\mathbf{n}}}{k_i}$ represents the cosine of the angle between the wave vector $\mathbf{k}_i$ and the line-of-sight direction $\hat{\mathbf{n}}$. After symmetrization, we obtain:
\begin{align}
Z_{2}(\mathbf{k}_{1},\mathbf{k}_{2})
& =\left(b_1-b_{\nabla^{2}\delta}k^{2}-b_{\psi}f\right)F_{2}(\mathbf{k}_1,\mathbf{k}_2)
+\left[(1-b_{\nabla^{2}\theta}k^{2})\mu^{2}+b_{\psi}\right]G_{2}(\mathbf{k}_1,\mathbf{k}_2)\nonumber \\
&+\left(\frac{1}{2}b_{2}+\frac{4}{21}b_{\psi}f\right)+\left(\frac{b_{s^2}}{2}-\frac{2}{7}b_{\psi}f\right)
S_{1}(\mathbf{k}_{1},\mathbf{k}_{2})\nonumber \\
&+\frac{1}{2}f_{2}k\mu\frac{\mu_{2}}{k_{2}}(1-b_{\nabla^{2}\theta}k^{2})\left[\left(b_1-b_{\nabla^{2}\delta}k^{2}\right)+f_{1}\mu_{1}^{2}(1-b_{\nabla^{2}\theta}k^{2})\right]\nonumber \\
&+\frac{1}{2}f_{1}k\mu\frac{\mu_{1}}{k_{1}}(1-b_{\nabla^{2}\theta}k^{2})\left[\left(b_1-b_{\nabla^{2}\delta}k^{2}\right)+f_{2}\mu_{2}^{2}(1-b_{\nabla^{2}\theta}k^{2})\right].
\label{eq: Z2_MG_sym}
\end{align}
Also according to Eq.~\eqref{eq:dgs},  the third order terms should be:
\begin{align}
\delta_{gs}^{(3)}(\mathbf{k})&= \delta_g^{(3)}(\mathbf{k})+\theta_g^{(3)}(\mathbf{k})\mu^{2}
\nonumber \\&+\int_{\mathbf{k}_{12}=\mathbf{k}}[\delta_g^{(2)}(\mathbf{k_{1}})+\theta_g^{(2)}(\mathbf{k_{1}})\mu_{1}^{2}]k\mu\frac{\mu_{2}}{k_{2}}\theta_g^{(1)}(\mathbf{k}_{2})\nonumber \\
&+\int_{\mathbf{k}_{12}=\mathbf{k}}[\delta_g^{(1)}(\mathbf{k_{1}})+\theta_g^{(1)}(\mathbf{k_{1}})\mu_{1}^{2}]k\mu\frac{\mu_{2}}{k_{2}}\theta_g^{(2)}(\mathbf{k}_{2})\nonumber \\
&+\int_{\mathbf{k}_{123}=\mathbf{k}}[\delta_g^{(1)}(\mathbf{k_{1}})+\theta_g^{(1)}(\mathbf{k_{1}})\mu_{1}^{2}]\frac{(k\mu)^{2}}{2}\frac{\mu_{2}}{k_{2}}\theta_g^{(1)}(\mathbf{k}_{2})\frac{\mu_{3}}{k_{3}}\theta_g^{(1)}(\mathbf{k}_{3})
\label{eq:dgs_3rd}
\end{align}

 We can also expand $\delta_{g}$ to the third order:%
\begin{align}
\delta_g^{(3)}(\mathbf{k}) &= ( b_{1}-b_{\nabla^{2}\delta}k^2-b_{\psi}f )\, \delta^{(3)}(\mathbf{k}) \nonumber\\
&\quad +b_{\psi}\theta^{(3)}(\mathbf{k})\nonumber\\
&\quad + (\frac{b_2}{2} +\frac{4}{21} b_{\psi} f)\int_{\mathbf{k}_{12}=\mathbf{k}} \delta^{(1)}(\mathbf{k}_1) \, \delta^{(2)}(\mathbf{k}_2) \nonumber\\
&\quad +( \frac{b_{s^2}}{2} -\frac{2}{7}b_{\psi}f)\int_{\mathbf{k}_{12}=\mathbf{k}} S_1(\mathbf{k}_1,\mathbf{k}_2) \, \delta^{(1)}(\mathbf{k}_1) \, \delta^{(2)}(\mathbf{k}_2) \nonumber\\ 
 & \quad+\frac{b_3}{6}\int_{\mathbf{k}_{123}=\mathbf{k}} \delta^{(1)}(\mathbf{k}_1) \, \delta^{(1)}(\mathbf{k}_2) \, \delta^{(1)}(\mathbf{k}_3)\nonumber\\
&\quad+\frac{b_{\delta s^2}}{2}\int_{\mathbf{k}_{123}=\mathbf{k}}S_1(\mathbf{k}_1,\mathbf{k}_{2}) \delta^{(1)}(\mathbf{k}_1) \, \delta^{(1)}(\mathbf{k}_2) \, \delta^{(1)}(\mathbf{k}_3)\nonumber\\
&\quad + b_{st} \int_{\mathbf{k}_{123}=\mathbf{k}} S_1(\mathbf{k}_1, \mathbf{k}_2 + \mathbf{k}_3) 
\left(  G_2(\mathbf{k}_2,\mathbf{k}_3)-f_2 F_2(\mathbf{k}_2,\mathbf{k}_3)\right) 
\delta^{(1)}(\mathbf{k}_1) \, \delta^{(1)}(\mathbf{k}_2)  \delta^{(1)}(\mathbf{k}_3)\nonumber\\
&\quad +\frac{b_{s^3}}{2}  \int_{\mathbf{k}_{123}=\mathbf{k}}  S_2(\mathbf{k}_1,\mathbf{k}_{1},\mathbf{k}_3) \delta^{(1)}(\mathbf{k}_2) \delta^{(1)}(\mathbf{k}_2)  \delta^{(1)}(\mathbf{k}_3)
\label{eq:delta_3rd_order_g}
\end{align}
The $b_{s^3}$ term contributes another kernel similar to $S_1(\mathbf{k}_1,\mathbf{k}_2)$, which is :%
\begin{align}
    S_2(\mathbf{k}_1,\mathbf{k}_2,\mathbf{k}_3) &=\frac{(\mathbf{k}_1\cdot\mathbf{k}_2)(\mathbf{k}_2\cdot\mathbf{k}_3)(\mathbf{k}_3\cdot\mathbf{k}_1)}{k_1^2 k_2^2 k_3^2} \nonumber\\
    &\quad -\frac{1}{3}\left[S_1(\mathbf{k}_1,\mathbf{k}_2)+S_1(\mathbf{k}_2,\mathbf{k}_3)+S_1(\mathbf{k}_3,\mathbf{k}_1)\right]-\frac{1}{9}.
    \label{eq:S_2kernel}
\end{align}
The first term in $S_2$ only depends on angles $\mu_{ij}$ and therefore contributes a constant term in the final expression for the power spectrum (see Eq.~\eqref{eq:P_oneloop} below) when integrated out.  Therefore, it can be absorbed into $Z_1$. Moreover,  the other terms in Eq.~\eqref{eq:S_2kernel}  can be  absorbed in $b_{\delta s^2}$ and $b_3$. As a consequence, we can safely ignore $b_{s^3}$.
Similarly, $b_3$ can be neglected since it gives a constant term in Eq.~\eqref{eq:P_oneloop} . So in the end we only have seven free bias parameters, namely $\{b_1,b_{\nabla^2\delta},b_2,b_{s^2},b_{\psi}\,b_{\delta s^2},b_{st}\}$.
Notice that now, since $f$  depends on $k$, combinations like $b_1-b_\psi f$ cannot be rewritten as a single bias function of time.
 Eq.~\eqref{eq:delta_3rd_order_g} should now be inserted into  $\delta_{gs}^{(3)}(\mathbf{k})$ in Eq.~\eqref{eq:dgs_3rd}.
Finally, we obtain the third-order kernel for biased tracers with the RSD effect:
\begin{align}
    Z_3(\mathbf{k}_{1},\mathbf{k}_{2},\mathbf{k}_{3}) &= ( b_{1}-b_{\nabla^{2}\delta}k^2-b_{\psi}f)F_{3}(\mathbf{k}_{1},\mathbf{k}_{2},\mathbf{k}_{3}) \nonumber \\
    &\quad + \left[(1-b_{\nabla^{2}\theta}k^{2})\mu^{2}+ b_{\psi}\right] G_{3}(\mathbf{k}_{1},\mathbf{k}_{2},\mathbf{k}_{3}) \nonumber \\
    &\quad + \left(\frac{b_2}{2} +\frac{4}{21} b_{\psi} f\right)F_2(\mathbf{k}_{2},\mathbf{k}_{3}) \nonumber \\
    &\quad + \left( \frac{b_{s^2}}{2} -\frac{2}{7} b_{\psi} f\right) S_1 (\mathbf{k}_1,\mathbf{k}_2) F_2 ( \mathbf{k}_2,\mathbf{k}_3)\nonumber \\
    &\quad + \frac{b_{\delta s^2}}{2}S_1(\mathbf{k}_{1},\mathbf{k}_{2})- b_{st} S_{1}(\mathbf{k}_{1},\mathbf{k}_{2}+\mathbf{k}_{3})\left(f_2F_{2}(\mathbf{k}_{2},\mathbf{k}_{3})-G_{2}(\mathbf{k}_{2},\mathbf{k}_{3})\right)\nonumber \\
      &\quad + k \mu f_3\frac{\mu_3}{k_3}(1-b_{\nabla^{2}\theta}k^2) \nonumber \\
      &\quad \quad \times \Bigg[\left[(1-b_{\nabla^{2}\theta}k^2)\mu_{12}^2+b_{\psi}\right] G_2(\mathbf{k}_1,\mathbf{k}_2)\nonumber \\
      &\quad\quad + ( b_{1}-b_{\nabla^{2}\delta}k^2-b_{\psi}f) F_2(\mathbf{k}_1,\mathbf{k}_2) \nonumber \\
      &\quad\quad+\left(\frac{b_2}{2}+\frac{4}{21}b_{\psi}f\right)+\left(\frac{b_{s^2}}{2}-\frac{2}{7}b_{\psi}f\right)S_1(\mathbf{k}_1,\mathbf{k}_2) \Bigg]\nonumber \\
      &\quad +\mu k\,(1-b_{\nabla^{2}\theta}k^2)\left[b_{1}-b_{\nabla^{2}\delta}k^2+(1-b_{\nabla^{2}\theta}k^{2})f_{3}\mu_{3}^{2}\right]\frac{\mu_{12}}{k_{12}}G_{2}(\mathbf{k}_{1},\mathbf{k}_{2})\nonumber \\
      &\quad +\frac{(k\mu)^2}{2}\frac{\mu_1}{k_1}\frac{\mu_2}{k_2}f_1 f_2 \left(1-b_{\nabla^{2}\theta}k^2\right)^2( b_{1}-b_{\nabla^{2}\delta}k^2+(1-b_{\nabla^{2}\theta}k^2)f_3\mu_3^2)
\end{align}
(to be symmetrized). 
These explicit and general expressions for $Z_2,Z_3$ for a $k$-dependent gravity represent the main results of the present work. 

\section{Power spectrum at one loop}
\label{sec: Pk at oneloop}
Let's collect the first three terms of the $\delta_{gs}$ (with the RSD and bias)
expansion obtained so far and write them more explicitly: 
\begin{align}
\delta_{gs}^{(1)}(\mathbf{k}) & =\delta^{(1)}(\mathbf{k})Z_{1}(\mathbf{k})\\
\delta_{gs}^{(2)}(\mathbf{k}) & =\int_{\mathbf{k}_{12}=\mathbf{k}}\delta^{(1)}(\mathbf{k}_{1})\delta^{(1)}(\mathbf{k}_{2})Z_{2}(\mathbf{k}_{1},\mathbf{k}_{2})\\
\delta_{gs}^{(3)}(\mathbf{k}) & =\int_{\mathbf{k}_{123}=\mathbf{k}}\delta^{(1)}(\mathbf{k}_{1})\delta^{(1)}(\mathbf{k}_{2})\delta^{(1)}(\mathbf{k}_{3})Z_{3}(\mathbf{k}_{1},\mathbf{k}_{2},\mathbf{k}_{3})\label{eq:fulldg}
\end{align}
where according to Eq.~\eqref{eq:lindg_1_order} 
\begin{equation}
Z_{1}(\mathbf{k})=b_1 - b_{\nabla^2\delta} k^2 +f\mu^2(1-b_{\nabla^2\theta} k^2)\, .
\end{equation}
Standard calculation \cite{Bernardeau_2002} shows that the one-loop spectrum for galaxies in redshift space is 
\begin{align}
P_{gg}(\mathbf{k},z) & =Z_1^2(\mathbf{k})P_{L}(\mathbf{k},z)+2P_{22}+6 Z_1(\mathbf{k})P_{13}(\mathbf{k},z)
\end{align}
where 
\begin{align}
\int P_{L}(k_{1})P_{L}(|\mathbf{k}-\mathbf{k}_{1}|)Z_{2}^{2}(\mathbf{k}_{1},\mathbf{k}-\mathbf{k}_{1})\frac{\mathrm{d}^{3}k_{1}}{(2\pi)^{3}} & \equiv P_{22} \label{eq:P_oneloop0}\\
P_{L}(k)\int P_{L}(k_{1})Z_{3}(\mathbf{k},\mathbf{k}_{1},-\mathbf{k}_{1})\frac{\mathrm{d}^{3}k_{1}}{(2\pi)^{3}} & \equiv P_{13}
\label{eq:P_oneloop}
\end{align}
To this spectrum, the usual UV corrections, the IR BAO damping, and shot noise should be
added (see e.g. \cite{Eisenstein_2007,2021arXiv211000016D,Ivanov:2018gjr}). Since
they are expected to be independent of the Horndeski kernels, we omit their expressions
here. 

\section{Conclusion}\label{sec:con}

In this work, we studied the nonlinear evolution of cosmological perturbations in theories with scale-dependent gravitational interactions, with a particular focus on Horndeski gravity. Using the fluid equations, we derived  expressions for the second-order and third-order kernels of Eulerian standard perturbation theory. These expressions are analytic up to a time integral, and they depend entirely only on the linear growth function, and on the  parameters of the functions $S,\mathcal{F}$. 
The final one-loop power spectrum also includes bias and redshift space distortion.
The formalism we developed is general and can be applied to any scenario where the linear growth function depends on scale.

As a proof-of-principle demonstration of our method, we consider Horndeski gravity and derive the expressions for the perturbative kernels when the $k$-dependent part is sub-dominant. We show that the nonlinear kernels can be fully expressed in terms of two time-dependent functions, \( h_1 \) and \( h_c \), which parametrize deviations from general relativity. This illustrates that the Wronskian method offers a practical framework for solving the growth equations and computing scale-dependent corrections to the perturbation kernels.

Our pipeline provides an alternative framework for calculating the one-loop galaxy power spectrum in scale-dependent theories. While this method is not necessarily expected to speed up calculations, it brings several advantages over the standard approach based on solving ordinary differential equations. 
First, it operates directly with the physical quantities such as the linear growth factor and the source function which enter directly in the equations of motion and Poisson equation.
Second, it reduces the problem to solving time-integrals, which is more numerically stable (e.g. using Gaussian quadrature method) than solving second-order differential equations on a grid. Clearly, one still has to perform the integrations for each set of cosmological parameters and, inside the kernel integrals (\ref{eq:P_oneloop0}) and (\ref{eq:P_oneloop}), over the ${\bf k_1}$ variables.
If the source $S(k,z)$ is separable, the overall $k$-dependence of the kernels can be factorized out (see App. \ref{app:sep}).
Thus, our method streamlines the calculation of the perturbative kernels within a physically motivated and numerically stable framework.

Our pipeline can be applied in several directions. 
First, we plan to perform a Fisher forecast for the precision of the cosmological measurements within scale-dependent modified gravity models.
Second, our framework allows for the evaluation of constraints in the presence of massive neutrinos by exploiting the accurate perturbative kernels. 
Third, our approach can be implemented in a fast code suitable for MCMC parameter estimation along the lines of~\cite{Rodriguez-Meza:2023rga}. 
We leave these research directions for future exploration. 


\vspace{0.5cm}
\noindent {\bf Acknowledgments }
\newline
\newline
ZZ acknowledges financial support from the German Research Foundation (DFG) under Germany’s Excellence Strategy – EXC 2181/1 – 390900948 (the Heidelberg STRUCTURES Excellence Cluster) for her research visits to Geneva, and thanks the University of Geneva for their hospitality. 
AC and MK acknowledge funding from the Swiss National Science Foundation.
LA acknowledges support by DFG  under Germany's Excellence Strategy EXC 2181/1 - 390900948 (the Heidelberg STRUCTURES Excellence Cluster) and under Project  554679582 "GeoGrav: Cosmological Geometry
and Gravity with nonlinear physics". LA acknowledges support  by Tamkeen under the NYU Abu Dhabi Research Institute grant CASS.


\appendix 

  \section{Relation with the $\alpha$-parametrization}
\label{sec:alpha}

In this Appendix we discuss the relation between the $h_i$ parameters introduced in Eq.~\eqref{eq: A_k_honedeski} and the popular $\alpha$-parametrization of Ref. \cite{Bellini:2014fua}.

The Horndeski scalar field gives rise to a Yukawa correction that  in Fourier space is given by 
\begin{equation}
    S(k,N)=\frac{3}{2}\Omega_m(N) Y(k,N),
\end{equation}
where
\begin{align}
 & \,Y=h_{1}\left(\frac{1+k^{2}h_{5}}{1+k^{2}h_{3}}\right)\,,\label{eq:etay-1}
\end{align}
(a similar form, in which two more functions $h_2,h_4$ enter, describes the effect of the Horndeski field on the anisotropic stress). An equivalent form is 
\begin{align}
 & \,Y=h_{1}\left(1+\frac{\alpha_{t}k^{2}}{m^{2}+k^{2}}\right)\, ,\label{eq:etay-alt}
\end{align}
where 
\begin{eqnarray}
\alpha_{t} & \equiv & (h_{5}-h_{3})/h_{3} \,  ,\\
m^{2} & \equiv & 1/h_{3} \,  .
\end{eqnarray}
Therefore we see that
\begin{equation}
    S_{kz}(k,N) = \frac{3}{2}\Omega_m(N) h_1 \frac{\alpha_t k^2}{m^2 + k^2}.
\end{equation}
In real space, the potential for a point particle of mass $M$ is 
\begin{equation}
\Psi(r)=-h_{1}\frac{G_{N}M}{r}(1+\alpha_{t}e^{-mr}).
\end{equation}
We see then that the parameter $\alpha_t$ represents the strength of the fifth force induced by the Horndeski scalar field,   $m^{-1}$ expresses the interaction range, and $h_1$ the time variation of the Newton constant.
The relation between the ``observable'' parameters $h_{i}$ that
enter the Yukawa correction and the ``physical'' parameters $\alpha_{K,B,M,T}$
is \cite{Amendola:2019laa}
\begin{eqnarray}
h_{1} & = & \frac{\alpha_{T}+1}{M_{\star}^{2}} \, ,\label{eq:h-alpha}\\
h_{3} & = & \frac{1}{2\mu^{2}}\left((2-\alpha_{B})\alpha_{1}+2\alpha_{2}\right) \, , \label{eq:h3}\\
h_{5} & = & \frac{1}{\mu^{2}}\left(\frac{\alpha_{M}+1}{\alpha_{T}+1}\alpha_{1}+\alpha_{2}\right) \, ,\label{eq:h5}
\end{eqnarray}
where $M_*$ is the time-dependent effective Planck mass, 
\begin{eqnarray}
\mu^{2} & \equiv & -3[2\xi^{2}+\xi'+\xi(3+\alpha_{M})]\alpha_{B}-3\xi\alpha_{2}
\end{eqnarray}
as well as 
\begin{eqnarray}
\alpha_{1} & \equiv & \alpha_{B}+\left(\alpha_{B}-2\right)\alpha_{T}+2\alpha_{M} \, ,\label{eq:an}\\
\alpha_{2} & \equiv & \alpha_{B}\xi+\alpha'_{B}-2\xi-3(1+w_{m})\tilde{\Omega}_{m}
\end{eqnarray}  
for $\xi=H'/H$ and $\tilde{\Omega}_{m}=\frac{\rho_{m}}{3M_{\star}^{2}H^{2}}$.
Here $\rho_m$ includes all the components beside the scalar field,
i.e. baryons, dark matter, neutrinos, radiation.

From the $h_{i}-\alpha_{i}$ relations we can derive the Yukawa strength
\begin{eqnarray}
\alpha_{t}=\frac{h_{5}-h_{3}}{h_{3}} & = & \frac{\alpha_{1}^{2}}{((2-\alpha_{B})\alpha_{1}+2\alpha_{2})\left(\alpha_{T}+1\right)} \, .\label{streng}
\end{eqnarray}
Just to provide an example, if $\alpha_B=\alpha_T=0$, the combination $h_1(h_5-h_3)$ is simply $\alpha_M^2/M_*^2\mu^2$. 

The simplest case of Horndeski Lagrangian is perhaps the Brans--Dicke model. In this model, the coupling between the scalar field $\phi$ and the Ricci scalar $R$ leads, in the Einstein frame, to a constant coupling $Q$ between $\phi$ and matter (see, e.g., Ref.~\cite{Tsujikawa_2008}). We can compute our parameter in the Brans--Dicke model and show that $\alpha_t$ indeed reduces to the constant coupling $Q$. In the Brans--Dicke model, we have $\alpha_M = -\alpha_B = \frac{\phi'}{\phi}$ and $\alpha_T=0$. Using these relation along with definitions Eqs.~\eqref{eq:h3} and \eqref{eq:h5}, we can derive $h_5 - h_3 = \frac{\alpha_M^2}{2 H^2 \mu^2}$. Furthermore, using the expressions for the Brans--Dicke model given in \cite{Amendola:2019laa},  $h_3 = \frac{3 + 2\omega}{2 \phi m_{\phi}^2}$, $\mu^2 = \frac{3 \alpha_M m_{\phi}^2 \phi'}{3 H^2}$, we find that $\frac{h_5 - h_3}{h_3} = \frac{1}{3 + 2\omega}$. Finally, using the relation between the coupling parameter and the Brans--Dicke parameter, $3 + 2\omega = \frac{1}{2Q^2}$ (\cite{Tsujikawa_2008}), we obtain
\begin{equation}
    \alpha_t = \frac{h_5 - h_3}{h_3} = 2Q^2 \,.
\end{equation}
Another example is the $f(R)$ model.
In this case $\alpha_B = -\alpha_M$ and $\alpha_T = \alpha_K=0$. Then the Yukawa strength becomes $\alpha_t = \frac{1}{3c_s^2}$, where $c_s$ is the sound speed of scalar field perturbations. For $c_s = 1$, this gives $\alpha_t = 1/3$.
For $m\gg k $, the ratio $S_{kz}/S_z\approx k^2/3m^2$ can be arbitrarily small. In this regime our perturbative approach is then justified.

\section{Numerical tests} \label{sec:numerical_test}

The results of this work are based on a first-order expansion in $h_c$. As an illustrative test of this assumption, we examine in this appendix the range of parameters for which our solution for $f_{kz}$ in Eq.~\eqref{eq:fk_solution} provides a good approximation to its exact numerical solution. Since we are mostly interested in the $k$-dependent part, we assume here $h_1=1$. The $k$-independent part of the growth, $f_z$, governed by Eq.~\eqref{eq:evolution_f_z}, is solved numerically.

\begin{figure}
    \centering
    \includegraphics[width=0.55\linewidth]{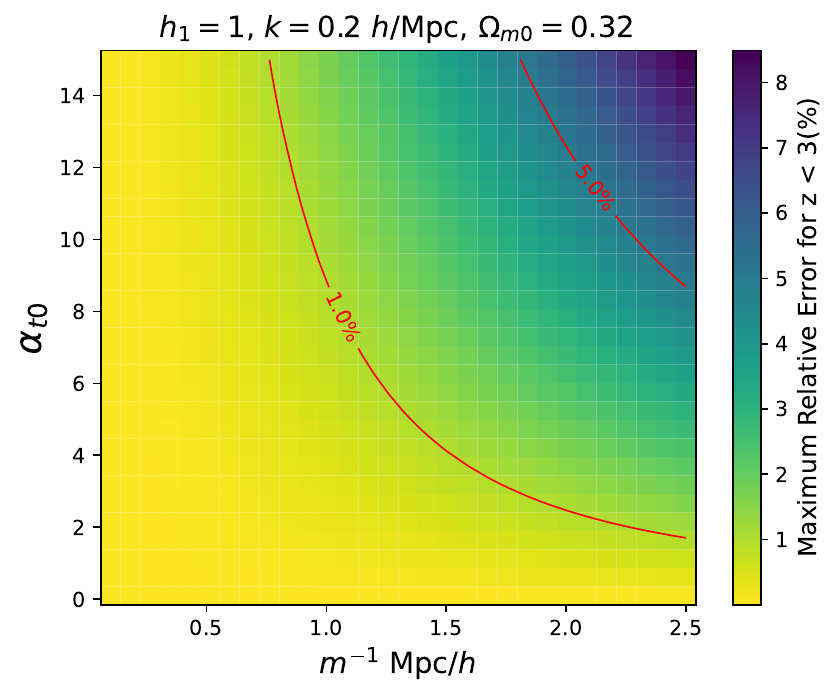}
    \caption{Maximum relative error of the $f_{kz}$  approximation for $z < 3$ in parameter space. The color indicates the magnitude of the maximum relative error between the exact numerical solution and our approximation, with fixed $h_{1}=1.0$, $k=0.2 h/\text{Mpc}$, $H_0 = 73.0 \, \text{km/s/Mpc}$ and $\Omega_{m0}=0.32$.}
    \label{fig:max_error_h30_alpha_t0_grid}
\end{figure}

\begin{figure}
    \centering
    \begin{minipage}{0.48\textwidth}
        \centering
        \includegraphics[width=\textwidth]{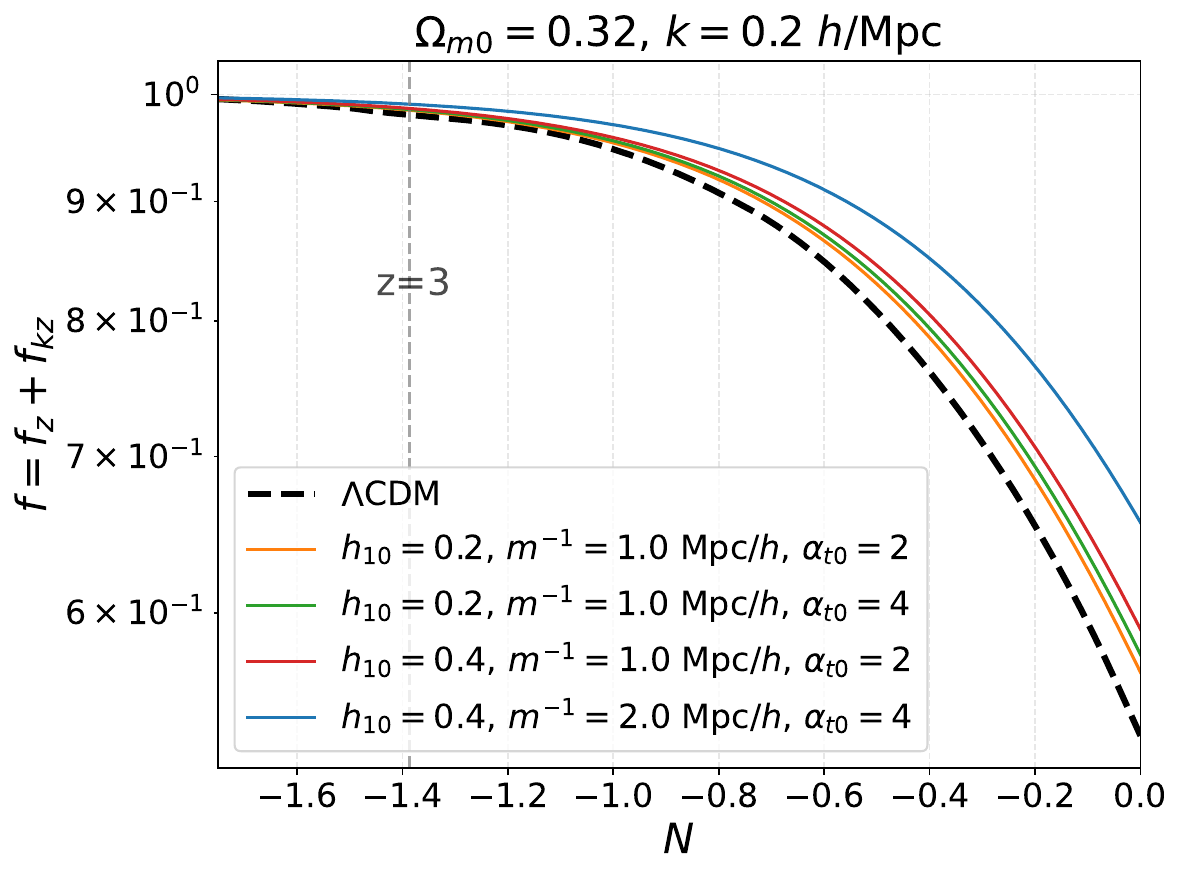}
    \end{minipage}%
    \hfill
    \begin{minipage}{0.48\textwidth}
        \centering
        \includegraphics[width=\textwidth]{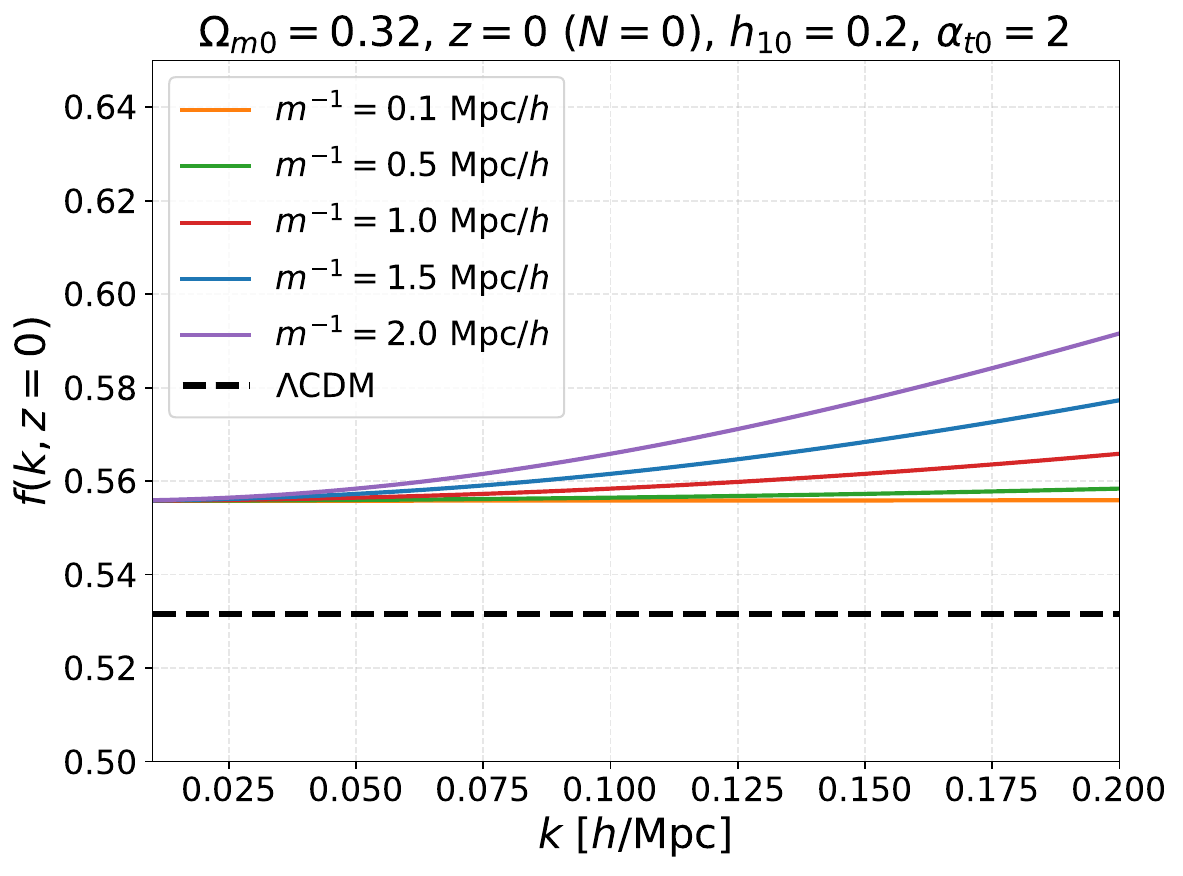}
    \end{minipage}
    \caption{\textbf{Left:} Comparison of the total growth rate \( f \) for various parameter choices, with fixed $H_0 = 73.0 \, \text{km/s/Mpc}$ and $\Omega_{m0} = 0.32$. The dashed vertical line indicates redshift $ z = 3$. \textbf{Right:} The growth rate at the current epoch as a function of $k$, with fixed $H_0 = 73.0 \, \text{km/s/Mpc}$, $h_{10}=0.2$, $\alpha_{t0}=2$ and $\Omega_{m0}=0.32$.
    }
    \label{fig:growth_rate}
\end{figure}

Several parameterizations of the Horndeski functions have been proposed in literature (see e.g. \cite{Bellini:2015xja,ade2016planck,kennedy2018reconstructing}) mostly based on simplicity and on the expectation that the modified gravity effects are associated to dark energy and therefore important only at late time. Here, we choose to parametrize $\alpha_t$ (the interaction coupling defined in App. \ref{sec:alpha}) in a similar way:
\begin{equation}
\alpha_t=\alpha_{t0}\Omega_\Lambda(N) \, ,
\label{eq:alpha_t_para}
\end{equation}
where $\Omega_\Lambda=(1-\Omega_{m0})H_0^2/H^2$, and $H$ is the usual $\Lambda$CDM function. Time-dependent couplings arise naturally in scalar-tensor models beyond Brans--Dicke. We assume instead for simplicity that $m$ is constant in time. We consider only $\alpha_{t0},m>0$ to ensure stability.
Then, the Horndeski parameters $h_3,h_5$ are
\begin{eqnarray}
h_3  &=& \frac{1}{m^2}= \text{constant} \,  ,\\
h_5 &=& h_3 \,(\alpha_{t}+1)=h_3 \,( \alpha_{t0}\, \Omega_{\Lambda}(N)+1) \, .
\end{eqnarray}
Notice that $h_3,h_5$ have dimensions of Mpc$^2/h^2$. We compare in Fig. \ref{fig:max_error_h30_alpha_t0_grid}
the maximum relative error varying our two free parameters, namely $\alpha_{t0}$ and $1/m$, within the observable redshift range $z<3$, between the numerical and analytical growth rate $f_{kz}$. We have fixed $k=0.2 h/$Mpc, which is approximately the highest wavenumber at which the nonlinear correction is still reliable. As expected, we find that for $1/m \lesssim 1$ Mpc$/h$, i.e. a short interaction range, the approximation is accurate for a very large region of $\alpha_{t0}$. Meanwhile, in Fig. \ref{fig:growth_rate}, left panel, we show the evolution of the growth rate over time for a range of different parameters. For similar reason as Eq.~\eqref{eq:alpha_t_para}, and to ensure $h_1=1$ at early times, we chose the parameterization of $h_1$ as
\begin{equation}
\label{h1_temp}
h_1=1+h_{10}\, \Omega_{\Lambda}(N) \, ,
\end{equation}
where $h_{10}$ is a dimensionless parameter. The scale dependence in this model is most significant at wavenumber $k > m$, as we show in the right panel of Fig. \ref{fig:growth_rate}. The large-scale limit of the $f(k,z=0)$ in Fig.~\ref{fig:growth_rate} (right panel) does not match the $\Lambda$CDM prediction due to  $h_{10} \ne 0$ in Eq.~\eqref{h1_temp}.

\section{Detailed derivations} \label{sec:detailed}
In this appendix, we provide detailed derivations of several equations used in the main text.

\textbf{Third-order continuity equation (Eq.~\eqref{eq:full_continuity}):}
\begin{align} 
\delta_{\textbf{k}}^{(3)}{}' - \theta^{(3)}_{\textbf{k}} & = \frac{1}{2} \int_{\textbf{k}_{12}=\textbf{k}} \left[ \alpha(\textbf{k}_{1},\textbf{k}_{2}) f_1 \delta^{(1)}_{\textbf{k}_{1}} \delta^{(2)}_{\textbf{k}_{2}} + \alpha(\textbf{k}_{2},\textbf{k}_{1}) f_2 \delta^{(1)}_{\textbf{k}_{2}} \delta^{(2)}_{\textbf{k}_{1}} \right] \nonumber \\
 &\quad + \frac{1}{2} \int_{\textbf{k}_{12}=\textbf{k}} \left[ \alpha(\textbf{k}_{1},\textbf{k}_{2}) \theta^{(2)}_{\textbf{k}_{1}} \delta^{(1)}_{\textbf{k}_{2}} + \alpha(\textbf{k}_{2},\textbf{k}_{1}) \theta^{(2)}_{\textbf{k}_{2}} \delta^{(1)}_{\textbf{k}_{1}} \right] \nonumber \\
&= \frac{1}{3} \left\{ \int_{\textbf{k}_1 + \textbf{q}_{23} = \textbf{k}} \alpha(\textbf{k}_1, \textbf{q}_{23}) f_1 \delta_{\textbf{k}_1} \delta_{\textbf{q}_{2}} \delta_{\textbf{q}_{3}} F_{2}(\textbf{q}_{2}, \textbf{q}_{3})  \right\}_{\text{cyc}} \nonumber \\
&\quad + \frac{1}{3}  \left\{ \int_{\textbf{q}_{13} + \textbf{k}_2 = \textbf{k}} \alpha(\textbf{q}_{13}, \textbf{k}_2) \delta_{\textbf{k}_2} \delta_{\textbf{q}_1} \delta_{\textbf{q}_3} G_{2}(\textbf{q}_1, \textbf{q}_3)  \right\}_{\text{cyc}} \nonumber \\
&= \frac{1}{3} \left\{ \int_{\textbf{k}_1 + \textbf{k}_{23} = \textbf{k}} \alpha_{1,23} f_1 \delta_{\textbf{k}_1} \delta_{\textbf{k}_2}\delta_{\textbf{k}_3} F_{2}(\textbf{k}_2, \textbf{k}_3) \right\}_{\text{cyc}} \nonumber \\
&\quad + \frac{1}{3} \left\{\int_{\textbf{k}_{13} + \textbf{k}_2 = \textbf{k}} \alpha_{13,2} \delta_{\textbf{k}_1} \delta_{\textbf{k}_2} \delta_{\textbf{k}_3} G_{2}(\textbf{k}_1, \textbf{k}_3) \right\}_{\text{cyc}} \nonumber \\
&= \frac{1}{3} \int_{\textbf{k}_{123} = \textbf{k}} \hat{\alpha}(\textbf{k}_{1}, \textbf{k}_{2}, \textbf{k}_{3}) \delta_{\textbf{k}_{1}} \delta_{\textbf{k}_{2}} \delta_{\textbf{k}_{3}}  \,, 
\end{align}

\textbf{Derivation of $R_{h_c^0}$ (Eq.~\eqref{eq:R_1z}):} %
In the derivation of \( R_{h_c^0} \), we decompose \( R \) into two terms according to different orders of \( h_c \), namely \( R = R_{h_c^0} + R_{h_c^1} \), by splitting
\( \hat{\alpha} = \hat{\alpha}_{h_c^0} + \hat{\alpha}_{h_c^1} \) and \( \hat{\beta} = \hat{\beta}_{h_c^0} + \hat{\beta}_{h_c^1} \). This results in
\begin{equation}
    R_{h_c^0}=  \frac{1}{3}\hat{\alpha}_{h_c^0}'+ \frac{2}{3}\hat{\beta}_{h_c^0}+\frac{1}{3}\hat{\alpha}_{h_c^0}(3f_z+\mathcal{F})\,,
\end{equation}
where
\begin{equation}
    \frac{1}{3}\hat{\alpha}_{h_c^0}' = \frac{1}{3}\left\{\alpha_{1,23} \Big[ f'_z\, F_{2, h_c^0}(\mathbf{k}_2, \mathbf{k}_3) + f_z\, F_{2,h_c^0}'(\mathbf{k}_2, \mathbf{k}_3) \Big]
 + \alpha_{23,1}\, G_{2,h_c^0}'(\mathbf{k}_2, \mathbf{k}_3)
\right\}_{\text{cyc}} \,.
\end{equation}
Using the explicit expressions for $\hat{\alpha}_{h_c^0}$, $\hat{\beta}_{h_c^0}$, $F_{2,h_c^0}$, and $G_{2,h_c^0}$ below in this appendix, we can obtain a more explicit expression for \( R_{h_c^0} \) in Eq.~\eqref{eq:R_1z}:
\begin{align} \label{eq: R_h_c^0}
    R_{h_c^0} & = \frac{1}{3}\left\{\alpha_{1,23} \Big[ f'_z\, F_{2, h_c^0}(\mathbf{k}_2, \mathbf{k}_3) + f_z\, F_{2,h_c^0}'(\mathbf{k}_2, \mathbf{k}_3) \Big]
 + \alpha_{23,1}\, G_{2,h_c^0}'(\mathbf{k}_2, \mathbf{k}_3)
\right\}_{\text{cyc}} \nonumber \\
 & + \frac{2}{3}\Big\{\beta_{1,23}f_zG_{2,h_c^0}(\mathbf{k}_{2},\mathbf{k}_{3}) \Big\}_{\text{cyc}} + \frac{1}{3}\Big\{\alpha_{1,23}f_zF_{2,h_c^0}(\mathbf{k}_{2},\mathbf{k}_{3})+\alpha_{12,3}G_{2,h_c^0}(\mathbf{k}_{1},\mathbf{k}_{2})\Big\}_{\text{cyc}} (3f_z+\mathcal{F}) \,.
\end{align}

\textbf{Explicit expressions for \( \hat{\alpha}_{h_c^0} \), \( \hat{\alpha}_{h_c^1} \), \( \hat{\beta}_{h_c^0} \), and \( \hat{\beta}_{h_c^1} \):} %
\begin{align}
\hat{\alpha}_{h_c^0}& =\Big\{\alpha_{1,23}f_zF_{2,h_c^0}(\mathbf{k}_{2},\mathbf{k}_{3})+\alpha_{12,3}G_{2,h_c^0}(\mathbf{k}_{1},\mathbf{k}_{2})\Big\}_{\text{cyc}}          \\
\hat{\alpha}_{h_c^1}& =  \Big\{\alpha_{1,23}f_{kz}(k_1)F_{2,h_c^0}(\mathbf{k}_{2},\mathbf{k}_{3})+\alpha_{1,23}f_{z}F_{2,h_c^1}(\mathbf{k}_{2},\mathbf{k}_{3})+\alpha_{12,3}G_{2,h_c^1}(\mathbf{k}_{1},\mathbf{k}_{2})\Big\}_{\text{cyc}}           \\
\hat{\beta}_{h_c^0}& = \Big\{\beta_{1,23}f_zG_{2,h_c^0}(\mathbf{k}_{2},\mathbf{k}_{3}) \Big\}_{\text{cyc}}       \\
\hat{\beta}_{h_c^1}& = \Big\{\beta_{1,23}f_{kz}(k_1)G_{2,h_c^0}(\mathbf{k}_{2},\mathbf{k}_{3})+\beta_{1,23}f_zG_{2,h_c^1}(\mathbf{k}_{2},\mathbf{k}_{3})\Big\}_{\text{cyc}}       
\end{align}
with
\begin{align}
F_{2,h_c^0} &= \frac{1}{2} + \frac{3\mathcal{A}_{h_c^0}}{14} 
+ \left( \frac{1}{2} - \frac{3\mathcal{B}_{h_c^0}}{14}  \right) \frac{(\mathbf{k}_1 \cdot \mathbf{k}_2)^2}{k_1^2 k_2^2} 
+ \frac{\mathbf{k}_1 \cdot \mathbf{k}_2}{2 k_1 k_2} \left( \frac{k_2}{k_1} + \frac{k_1}{k_2} \right) \,\label{eq:F2_hc0}  \\
F_{2,h_c^1} &= \frac{3 \mathcal{A}_{h_c^1}}{14} 
 - \frac{3\mathcal{B}_{h_c^1}}{14}  \frac{(\mathbf{k}_1 \cdot \mathbf{k}_2)^2}{k_1^2 k_2^2}  \, \label{eq:F2_hc1}\\
G_{2,h_c^0} &= \frac{3\mathcal{A}_{h_c^0}f_z}{7} + \frac{3\mathcal{A}_{h_c^0}'}{14}
+ \left( f_z - \frac{3\mathcal{B}_{h_c^0}f_z}{7} - \frac{3\mathcal{B}'_{h_c^0}}{14} \right) \frac{(\mathbf{k}_1 \cdot \mathbf{k}_2)^2}{k_1^2 k_2^2}
+ \frac{\mathbf{k}_1 \cdot \mathbf{k}_2}{2 k_1 k_2} \left( \frac{f_z k_2}{k_1} + \frac{f_z k_1}{k_2} \right) \label{eq:G2_hc0}\\ 
G_{2,h_c^1} &= \frac{3\mathcal{A}_{h_c^1}f_z}{7} + \frac{3\mathcal{A}_{h_c^0}\left(f_{kz}(k_1)+f_{kz}(k_2)\right)}{14} + \frac{3\mathcal{A}'_{h_c^1}}{14}  \nonumber  \\
&+ \left[ \frac{f_{kz}(k_1) + f_{kz}(k_2)}{2} -\frac{3\mathcal{B}_{h_c^1}f_z}{7} - \frac{3\mathcal{B}_{h_c^0}\left(f_{kz}(k_1)+f_{kz}(k_2)\right)}{14} - \frac{3\mathcal{B}'_{h_c^1}}{14} \right] \frac{(\mathbf{k}_1 \cdot \mathbf{k}_2)^2}{k_1^2 k_2^2}  \nonumber \\ \,
&+ \frac{\mathbf{k}_1 \cdot \mathbf{k}_2}{2 k_1 k_2} \left[ \frac{f_{kz}(k_2) k_2}{k_1} + \frac{f_{kz}(k_1) k_1}{k_2} \right] \,,     \label{eq:G2_hc1}\,
\end{align}
and 
\begin{align}
\mathcal{A}_{h_c^0} & = \mathcal{B}_{h_c^0}  =\frac{7D_{12,z}}{3D_{+}^{2}}   \\
\mathcal{A}_{h_c^1} &  =\frac{7D_{12,z}}{3D_{+}^{2}}\left[\int_{N_{0}}^{N}\mathrm{d}x\frac{D_{-}D_{+} \big(S_{kz}(k_1)+S_{kz}(k_2)\big)}{W} - \frac{D_{-}}{D_{+}}\int_{N_{0}}^{N}\mathrm{d}x\frac{D_{+}^{2} \big(S_{kz}(k_1)+S_{kz}(k_2)\big)}{W}\right]+ \nonumber \\
 & \frac{7}{3}\left\{ -\frac{1}{D_{+}}\int_{N_{0}}^{N}\mathrm{d}x\frac{D_{-}\hat{\mathcal{I}}_{\mathcal{A}}}{W}+\frac{D_{-}}{D_{+}^{2}}\int_{N_{0}}^{N}\mathrm{d}x\frac{D_{+}\hat{\mathcal{I}}_{\mathcal{A}}}{W}\right\}   \,      \\  
 \mathcal{B}_{h_c^1} & =\frac{7D_{12,z}}{3D_{+}^{2}}\left[\int_{N_{0}}^{N}\mathrm{d}x\frac{D_{-}D_{+} \big(S_{kz}(k_1)+S_{kz}(k_2)\big)}{W} - \frac{D_{-}}{D_{+}}\int_{N_{0}}^{N}\mathrm{d}x\frac{D_{+}^{2} \big(S_{kz}(k_1)+S_{kz}(k_2)\big)}{W}\right]+ \nonumber\\
 &  \frac{7}{3}\left\{ -\frac{1}{D_{+}}\int_{N_{0}}^{N}\mathrm{d}x\frac{D_{-}\hat{\mathcal{I}}_{\mathcal{B}}}{W}+\frac{D_{-}}{D_{+}^{2}}\int_{N_{0}}^{N}\mathrm{d}x\frac{D_{+}\hat{\mathcal{I}}_{\mathcal{B}}}{W}\right\} 
\end{align}

\section{Kernels in the $k$-independent and EdS limits}
In this appendix, we demonstrate that the kernels derived in Sec.~\ref{sec: kernels} reduce to the standard result in certain limiting cases. We first consider the $k$-independent limit of the growth functions, followed by the Einstein-de Sitter (EdS) limit, for both the second- and third-order kernels. 

\subsection{Second-order kernels}
\textbf{k-independent limit:} 
The second-order growth equation for $D_{12,\mathcal{A}}$ in this limit reduces to: 
\begin{equation}
D_{12, \mathcal{A}_z}^{\prime\prime} + \mathcal{F}D_{12,\mathcal{A}_z}^{\prime} - S_z D_{12,\mathcal{A}_z} = S_zD_{z}^2 \,,
\label{eq:D_12_AB_k_ind}
\end{equation}
 and the equation for $D_{12,\mathcal{B}}$ takes the same form. Therefore, we have $D_{12,\mathcal{A}} = D_{12,\mathcal{B}}$ and consequently $\mathcal{A} = \mathcal{B}$, both only depend on time. Thus, the second-order kernels, Eqs.~\eqref{eq:F2kernel} and~\eqref{eq:F_2_G_2_kernel}, can be written as follows:
\begin{align}
F_{2}(\mathbf{k_{1}},\mathbf{k_{2}}) &= \left(\frac{1}{2}+\frac{3}{14}\mathcal{A}\right)\tilde{\alpha}_{1,2} + \left(\frac{1}{2}-\frac{3}{14}\mathcal{A}\right)\beta_{1,2} \, , \label{eq: k_inde_F2}\\
G_{2}(\mathbf{k_{1}},\mathbf{k_{2}}) &= \frac{3\mathcal{A}^\prime+6f_z\mathcal{A}}{14}\tilde{\alpha}_{1,2} + \left(f_z-\frac{3\mathcal{A}^\prime+6f_z\mathcal{A}}{14}\right)\beta_{1,2} \label{eq: k_inde_G2}\,,
\end{align}
where \( \tilde{\alpha}_{1,2} \equiv \frac{1}{2}(\alpha_{1,2} + \alpha_{2,1}) \). From Eq.~\eqref{eq: kernels}, the second-order density perturbation $\delta^{(2)}$ is given by
\begin{equation}
    \delta^{(2)}(\mathbf{k}, N) = D_z^2 \int_{\mathbf{k}_{12} = \mathbf{k}} F_2(\mathbf{k}_1, \mathbf{k}_2; N)\, \delta_{0}(\textbf{k}_{1})\, \delta_{0}(\textbf{k}_{2}) \,.
\end{equation}

We can express $\delta^{(2)}$  as a linear combination of separable contributions,
\begin{equation}
\delta^{(2)} = g_{2,A}(N)\,A(k) \;+\; g_{2,B}(N)\,B(k)
\end{equation}
with
\begin{align}
A(k) &= \frac{5}{7}\int_{\mathbf{k}_{12}=\mathbf{k}} \, \tilde{\alpha}_{1,2}  \,\delta_{0}(\textbf{k}_{1})\, \delta_{0}(\textbf{k}_{2}) 
\,, \\
B(k) &= \frac{2}{7}\int_{\mathbf{k}_{12}=\mathbf{k}} \, \beta_{1,2}  \, \delta_{0}(\textbf{k}_{1})\, \delta_{0}(\textbf{k}_{2}) \,.
\end{align} %
Comparing the above equations, we identify $g_{2,A}(N) \equiv \frac{7}{5}D_z^2(\frac{1}{2} + \frac{3}{14}\mathcal{A})$ and $g_{2,B}(N) \equiv \frac{7}{2} D_z^2(\frac{1}{2} - \frac{3}{14}\mathcal{A})$, which serve as the second-order growth factor associated with the mode-coupling terms $A(k)$ and $B(k)$. Combining these two relations with Eq.~\eqref{eq:mathAB-1}, we obtain
\begin{equation}
    D_{12, \mathcal{A}_z} = \frac{10}{7} g_{2,A} - D_z^2 = D_z^2-\frac{4}{7}g_{2,B}\,.
\end{equation}
Combining Eqs.~\eqref{eq:D_12_AB_k_ind}  and ~\eqref{eq:Dz-1-2}, we obtain explicit expressions for $g_{2,A}$ and $g_{2,B}$
\begin{align}
    g_{2,A}^{\prime\prime}+\mathcal{F}g_{2,A}^{\prime}-S_{z}g_{2,A} &= \frac{7}{5}D_z^2(f_z^2+S_{z}) \,, \\
    g_{2,B}^{\prime\prime}+\mathcal{F}g_{2,B}^{\prime}-S_{z}g_{2,B} &= \frac{7}{2}D_z^2f_z^2 \,.
\label{eq:D2_k_ind}
\end{align}
Lastly, defining $\tilde{g}_{2A} \equiv g_{2A}/a^2, \tilde{g}_{2B} \equiv g_{2B}/a^2$, and using $S_z=3\Omega_m(a)/2$ and $\mathcal{F}=2+H'/H$, we recover the standard result, as found in, e.g., Ref. \cite{takahashi2008third}.  
 
\textbf{EdS limit:} in the EdS limit, we consider a flat, matter-dominated universe with $\Omega_m = 1$, $f=1$, $\mathcal{F}=1/2$ , $  S = 3/2$ and $ D_+ \propto a$. In this case, the mode-coupling functions reduce to $\mathcal{A}=\mathcal{B}=1$, and the second-order kernels take the well-known EdS form:
\begin{align}
F_{2,{\rm EdS}}(\mathbf{k_{1}},\mathbf{k_{2}})&=\frac{5}{7}+\frac{2(\mathbf{k_{1}}\cdot\mathbf{k_{2}})^{2}}{7k_{1}^{2}k_{2}^{2}}+\frac{\mathbf{k_{1}}\cdot\mathbf{k_{2}}}{2k_{1}k_{2}}\left(\frac{k_{2}}{k_{1}}+\frac{k_{1}}{k_{2}}\right)
\\
G_{2,{\rm EdS}}(\mathbf{k_{1}},\mathbf{k_{2}})&=\frac{3}{7}+\frac{4}{7}\frac{(\mathbf{k_{1}}\cdot\mathbf{k_{2}})^{2}}{k_{1}^{2}k_{2}^{2}}+\frac{\mathbf{k_{1}}\cdot\mathbf{k_{2}}}{2k_{1}k_{2}}\left(\frac{k_{2}}{k_{1}}+\frac{k_{1}}{k_{2}}\right) 
\,,
\end{align}
and Eq.~\eqref{eq:d12z} simplifies to:
\begin{equation}
D_{12,z}''+\frac{1}{2}D_{12,z}'-\frac{3}{2}D_{12,z}=\frac{3}{2}D_{+}^{2} \,,
\end{equation}
The homogeneous equation admits a growing mode $D_{+}= e^N$ (normalized to unity at the present epoch) and a decaying mode $D_{-}\propto e^{-3N/2}$. Substituting the Wronskian $W\propto -\frac{5}{2}e^{-N/2}$ into Eq.~\eqref{eq:D_12_z_Wronskian}, we obtain
\begin{align}
D_{12,z}= \frac{3}{5}e^N(e^N-e^{N_0}) - \frac{3}{5}e^{-3N/2}(\frac{2}{7}e^{7N/2}-\frac{2}{7}e^{7N_0/2}) \,.
\end{align}
Taking the limit $N_0\to-\infty$, we recover $D_{12,z}=\frac{3}{7}e^{2N}=\frac{3}{7}a^2$, as expected.

\subsection{Third-order kernels}
\textbf{k-independent limit:}  The third-order growth equation for $D_{123}$ in this limit becomes
\begin{equation}
D_{123,z}''+\mathcal{F}D_{123,z}'-S_zD_{123,z}=R_{h_c^0}6D_{z}^{3} \, .
\end{equation}
Combining Eqs. \eqref{eq:evolution_f_z}, \eqref{eq:2nd_F_2}, and \eqref{eq:2nd_G_2}, we obtain
\begin{align}
    R_{h_c^0} &= \frac{1}{3}\left\{ \tilde{\alpha}_{1,23}S_zF_{2}(\mathbf{k}_2, \mathbf{k}_3) + \tilde{\alpha}_{23,1}S_zF_{2}(\mathbf{k}_2, \mathbf{k}_3) + \tilde{\alpha}_{1,23}f_zG_{2}(\mathbf{k}_2, \mathbf{k}_3) + \tilde{\alpha}_{23,1}f_zG_{2}(\mathbf{k}_2, \mathbf{k}_3) \right\}_{\text{cyc}}  \nonumber\\ 
    & + \frac{1}{3}\left\{ \tilde{\alpha}_{1,23}\tilde{\alpha}_{2,3}f_z^2 + \tilde{\alpha}_{23,1}\beta_{2,3}f_z^2 + 2\beta_{1,23}f_zG_{2}(\mathbf{k}_2, \mathbf{k}_3) \right\}_{\text{cyc}} \, .
\end{align} %
Further employing Eqs. \eqref{eq: k_inde_F2}, and \eqref{eq: k_inde_G2}, we found the expression for \( R_{h_c^0} \) consists of six distinct terms, each written as a product of a time-dependent coefficient and a scale-dependent kernel contraction. For clarity, we present these terms in Table~\ref{tab:Rhc0_split}, where the time-dependent parts involve functions such as \( S_z \), \( f_z \), \( \mathcal{A} \), and \( \mathcal{A}' \), while the scale-dependent parts correspond to specific combinations of mode-coupling kernels, symmetrized over cyclic permutations of the wavevectors.

\begin{table}[ht]
    \centering
    \renewcommand{\arraystretch}{1.5}
    \begin{tabular}{|c|c|c|}
        \hline
        Term & Time-dependent coefficients & Scale-dependent term \\
        \hline
        1 & 
        $\displaystyle \frac{1}{3}\left[ S_z\left(\frac{1}{2} + \frac{3}{14}\mathcal{A} \right) 
        + f_z\frac{3\mathcal{A}' + 6f_z\mathcal{A}}{14} + f_z^2 \right]$ 
        & 
        $\left\{ \tilde{\alpha}_{1,23} \tilde{\alpha}_{2,3} \right\}_{\text{cyc}}$ \\
        \hline
        2 & 
        $\displaystyle \frac{1}{3}\left[ S_z\left(\frac{1}{2} + \frac{3}{14}\mathcal{A} \right) 
        + f_z\frac{3\mathcal{A}' + 6f_z\mathcal{A}}{14} \right]$ 
        & 
        $\left\{ \tilde{\alpha}_{23,1} \tilde{\alpha}_{2,3} \right\}_{\text{cyc}}$ \\
        \hline
        3 & 
        $\displaystyle \frac{1}{3}\left[ S_z\left(\frac{1}{2} - \frac{3}{14}\mathcal{A} \right)
        + f_z\left(f_z - \frac{3\mathcal{A}' + 6f_z\mathcal{A}}{14} \right) \right]$ 
        & 
        $\left\{ \tilde{\alpha}_{1,23} \beta_{2,3} \right\}_{\text{cyc}}$ \\
        \hline
        4 & 
        $\displaystyle \frac{1}{3}\left[ S_z\left(\frac{1}{2} - \frac{3}{14}\mathcal{A} \right)
        + f_z\left(f_z - \frac{3\mathcal{A}' + 6f_z\mathcal{A}}{14} \right) + f_z^2 \right]$ 
        & 
        $\left\{ \tilde{\alpha}_{23,1} \beta_{2,3} \right\}_{\text{cyc}}$ \\
        \hline
        5 & 
        $\displaystyle \frac{2}{3}f_z\frac{3\mathcal{A}' + 6f_z\mathcal{A}}{14}$ 
        & 
        $\left\{ \beta_{1,23} \tilde{\alpha}_{2,3} \right\}_{\text{cyc}}$ \\
        \hline
        6 & 
        $\displaystyle \frac{2}{3}f_z\left(f_z - \frac{3\mathcal{A}' + 6f_z\mathcal{A}}{14} \right)$ 
        & 
        $\left\{ \beta_{1,23} \beta_{2,3} \right\}_{\text{cyc}}$ \\
        \hline
    \end{tabular}
    \caption{Summary of the six terms contributing to $R_{h_c^0}$, with time- and scale-dependent components separated.}
    \label{tab:Rhc0_split}
\end{table}

Therefore $D_{123,z}$ can be decomposed as
\begin{equation}
    D_{123,z} = g_{3,A}(N)A_3(k) +   g_{3,\tilde{A}}(N)\tilde{A}_3(k) + g_{3,B}(N)B_3(k) + g_{3,\tilde{B}}(N)
    \tilde{B}_3(k) +  g_{3,C}(N)C_3(k) +  g_{3,D}(N)D_3(k) \, .
\end{equation}
From Eqs.~\eqref{eq: kernels} and 
\eqref{eq:final_F3}, the third-order density perturbation $\delta^{(3)}$ can be written as follows
\begin{align}
    \delta^{(3)}(\mathbf{k}, N) &= D_z^3 \int_{\mathbf{k}_{123} = \mathbf{k}} F_3(\mathbf{k}_1, \mathbf{k}_2, \mathbf{k}_3; N)\, \delta_{0}(\textbf{k}_{1})\, \delta_{0}(\textbf{k}_{2}) \,\delta_{0}(\textbf{k}_{2}) \nonumber \\
    &= \frac{1}{6} \int_{\mathbf{k}_{123} = \mathbf{k}}D_{123,z}(\mathbf{k}_1, \mathbf{k}_2, \mathbf{k}_3; N)\, \delta_{0}(\textbf{k}_{1})\, \delta_{0}(\textbf{k}_{2}) \,\delta_{0}(\textbf{k}_{3}) \nonumber \\
    &= g_{3,A}(N)I_A(k) +  g_{3,
    \tilde{A}}(N)I_{\tilde{A}}(k) + g_{3,B}(N)I_B(k) + g_{3,
    \tilde{B}}(N)I_{\tilde{B}}(k) + g_{3,C}(N)I_C(k) +  g_{3,D}(N)I_D(k) 
    \,,
\end{align}
where
\begin{align}
    I_A(k) & = \frac{1}{6} \int_{\mathbf{k}_{123}=\mathbf{k}} \, \left\{\tilde{\alpha}_{1,23}\tilde{\alpha}_{2,3}  \right\}_{\text{cyc}}\delta_{0}(\textbf{k}_{1})\, \delta_{0}(\textbf{k}_{2}) \,\delta_{0}(\textbf{k}_{3})     \,  \\
    I_{\tilde{A}}(k) & = \frac{1}{6}  \int_{\mathbf{k}_{123}=\mathbf{k}} \, \left\{\tilde{\alpha}_{23,1}\tilde{\alpha}_{2,3}  \right\}_{\text{cyc}}\delta_{0}(\textbf{k}_{1})\, \delta_{0}(\textbf{k}_{2}) \,\delta_{0}(\textbf{k}_{3}) \,  \\
    I_B(k) & = \frac{1}{6}  \int_{\mathbf{k}_{123}=\mathbf{k}} \, \left\{\tilde{\alpha}_{1,23}\tilde{\beta}_{2,3}  \right\}_{\text{cyc}}\delta_{0}(\textbf{k}_{1})\, \delta_{0}(\textbf{k}_{2}) \,\delta_{0}(\textbf{k}_{3})   \,  \\
    I_{\tilde{B}}(k) & = \frac{1}{6}  \int_{\mathbf{k}_{123}=\mathbf{k}} \, \left\{\tilde{\alpha}_{23,1}\tilde{\beta}_{2,3}  \right\}_{\text{cyc}}\delta_{0}(\textbf{k}_{1})\, \delta_{0}(\textbf{k}_{2}) \,\delta_{0}(\textbf{k}_{3}) \,  \\
    I_C(k) & = \frac{1}{6}  \int_{\mathbf{k}_{123}=\mathbf{k}} \, \left\{\tilde{\beta}_{1,23}\tilde{\alpha}_{2,3}  \right\}_{\text{cyc}}\delta_{0}(\textbf{k}_{1})\, \delta_{0}(\textbf{k}_{2}) \,\delta_{0}(\textbf{k}_{3}) \,  \\
    I_D(k) & =  \frac{1}{6}  \int_{\mathbf{k}_{123}=\mathbf{k}} \, \left\{\tilde{\beta}_{1,23}\tilde{\beta}_{2,3}  \right\}_{\text{cyc}}\delta_{0}(\textbf{k}_{1})\, \delta_{0}(\textbf{k}_{2}) \,\delta_{0}(\textbf{k}_{3})    \,.
\end{align}
As pointed out in ~\cite{takahashi2008third}, the evolution of $g_{3,\tilde{A}}$ and $g_{3,\tilde{B}}$ depend fully on $g_{3, A}, g_{3, B}, g_{3, C}$ and $g_{3, D}$, while the four independent $g$-functions obey
\begin{align}
 g_{3,A}^{\prime\prime}+\mathcal{F}g_{3,A}^{\prime}-S_{z}g_{3,A} &= 2D_z^3\Bigg[S_z\left(\frac{1}{2}+\frac{3}{14}\mathcal{A}\right) + f_z\frac{3\mathcal{A}^\prime+6f_z\mathcal{A}}{14} +f_z^2\Bigg] \,, \\
 g_{3,B}^{\prime\prime}+\mathcal{F}g_{3,B}^{\prime}-S_{z}g_{3,B} &=  2D_z^3\Bigg[S_z\left(\frac{1}{2}-\frac{3}{14}\mathcal{A}\right)+f_z\left(f_z-\frac{3\mathcal{A}^\prime+6f_z\mathcal{A}}{14}\right)\Bigg]\,,     \\
 g_{3,C}^{\prime\prime}+\mathcal{F}g_{3,C}^{\prime}-S_{z}g_{3,C} &= 4D_z^3f_z\frac{3\mathcal{A}^\prime+6f_z\mathcal{A}}{14}\,, \\
 g_{3,D}^{\prime\prime}+\mathcal{F}g_{3,D}^{\prime}-S_{z}g_{3,D} &= 4D_z^3f_z\left(f_z-\frac{3\mathcal{A}^\prime+6f_z\mathcal{A}}{14}\right) \,,  
\end{align}
which coincide with the equations found in~\cite{takahashi2008third}.

\textbf{EdS limit:} The evolution equation for $D_{123}$ in the EdS scenario becomes,
\begin{equation}
    D_{123}''+ \frac{1}{2}D_{123}'-\frac{3}{2}D_{123} = e^{3N}\left(4\hat{\beta}+7\hat{\alpha}\right) 
    \,,
\end{equation}
note that $\hat{\alpha}$ and $\hat{\beta}$ in this case do not depend on time. Thus, the solution for $D_{123}(\textbf{k}_{1},\textbf{k}_{2}, \textbf{k}_{3},N)$ is given by:
\begin{align}
D_{123}(\textbf{k}_{1},\textbf{k}_{2},\textbf{k}_{3},N)&=\left(4\hat{\beta}+7\hat{\alpha}\right)\left[-e^N\int_{N_{0}}^{N}\mathrm{d}x\frac{e^{-\frac{3}{2}x}\cdot e^{3x}}{-\frac{5}{2}e^{-x/2}}+e^{-\frac{3}{2}N}\int_{N_{0}}^{N}\mathrm{d}x\frac{e^x\cdot e^{3x}}{-\frac{5}{2}e^{-x/2}}\right]   \\
& =\left(4\hat{\beta
}+7\hat{\alpha}\right) \left[\frac{2}{5}e^N\int_{N_{0}}^{N}e^{2x}\mathrm{d}x -\frac{2}{5} e^{-\frac{3}{2}N}\int_{N_{0}}^{N}e^{\frac{9}{2}x}\mathrm{d}x\right]  \,.
\end{align}
Taking again the limit $N_0\to-\infty$, we obtain
\begin{equation}
    D_{123} = \left(4\hat{\beta}+7\hat{\alpha}\right)\left[
    \frac{1}{5}e^{3N}-\frac{4}{45}e^{3N}\right] = \frac{1}{9}\left(4\hat{\beta}+7\hat{\alpha}\right) e^{3N}; \quad \mathcal{A}_3 = \frac{7D_{123}}{3e^{3N}}= \frac{7}{27}\left(4\hat{\beta}+7\hat{\alpha}\right)
    \,.
\end{equation}
The third-order kernels derived in Eqs.~\eqref{eq:final_F3} and \eqref{eq:final_G3} then reduce to the EdS form, namely
\begin{align}
    F_{3,{\rm EdS}}(\mathbf{k_{1}},\mathbf{k_{2}},\mathbf{k_{3}}) & = \frac{1}{54}\left(4\hat{\beta}+7\hat{\alpha}\right) \nonumber \\
    & = \frac{1}{54}\left[4\Big\{\beta_{1,23}G_{2}(\mathbf{k}_{2},\mathbf{k}_{3}) \Big\}_{\text{cyc}} + 7\Big\{\alpha_{1,23}F_{2}(\mathbf{k}_{2},\mathbf{k}_{3})+\alpha_{13,2}G_{2}(\mathbf{k}_{1},\mathbf{k}_{3})\Big\}_{\text{cyc}}\right] \,, \\
    G_{3,{\rm EdS}}(\mathbf{k_{1}},\mathbf{k_{2}},\mathbf{k_{3}}) & = \frac{1}{18}\left(4\hat{\beta}+\hat{\alpha}\right) \nonumber \\
    & = \frac{1}{18}\left[4\Big\{\beta_{1,23}G_{2}(\mathbf{k}_{2},\mathbf{k}_{3}) \Big\}_{\text{cyc}} + \Big\{\alpha_{1,23}F_{2}(\mathbf{k}_{2},\mathbf{k}_{3})+\alpha_{13,2}G_{2}(\mathbf{k}_{1},\mathbf{k}_{3})\Big\}_{\text{cyc}}\right] \,.
\end{align}
\section{Separable case}
\label{app:sep}

In this section we present the main results in the case of a separable source function.

\subsection{Second-order kernels}
When the source function is separable, i.e. for 
\begin{align*}
    S_{kz}\equiv \mathcal{T}(N)  \mathcal{K}(k),
\end{align*}
the computation of the kernels becomes much faster, since the $k$-dependent part can be factorized out  of the integrals. In our Horndeski case, the separation occurs if $h_3=$const. Then,
 the time-dependent factor is   $\mathcal{T}(N)=S_z\,(h_5-h_3)$ and the scale-dependent factor takes the form $\mathcal{K}(k)= \frac{k^{2}}{1+h_{3}k^{2}}$.
It is noteworthy that this approach is valid under the assumption  $h_3 =m^{-2}$  remains constant. The scale-dependent correction $D_{kz}$ in Eq.~\eqref{eq: D_kz_withW} can be reformulated as,
\begin{align}
      D_{kz}&\equiv\mathcal{K}(k)\,\mathcal{X}_0(N)\\
    &  =\mathcal{K}\left[- D_{+}\int_{N_{0}}^{N}\mathrm{d}x\frac{D_{-}D_{+} \mathcal{T}}{W}+ D_{-}\int_{N_{0}}^{N}\mathrm{d}x\frac{D_{+}^{2} \mathcal{T}}{W}\right].
\end{align}
To facilitate numerical computation of $F_2$ and $G_2$ kernels in Eq. ~\eqref{eq:F_2_G_2_kernel}, we introduce a set of auxiliary functions that enable efficient pre-calculation. The time-dependent functions are defined as:
\begin{align}
\mathcal{X}_{1}(N)&\equiv -\frac{1}{D_{+}}\int_{N_{0}}^{N}\mathrm{d}x\frac{D_{-}D_{+}S_{z}\mathcal{X}_0}{W}
+\frac{D_{-}}{D_{+}^{2}}\int_{N_{0}}^{N}\mathrm{d}x\frac{D_{+}^{2}S_{z}\mathcal{X}_0}{W},\\
\mathcal{X}_{2}(N)&\equiv -\frac{1}{D_{+}}\int_{N_{0}}^{N}\mathrm{d}x\frac{D_{-}D_{12,z}\mathcal{T}}{W}
+\frac{D_{-}}{D_{+}^{2}}\int_{N_{0}}^{N}\mathrm{d}x\frac{D_{+} D_{12,z}\mathcal{T}}{W},\\
\mathcal{X}_{3}(N)&\equiv -\frac{1}{D_{+}}\int_{N_{0}}^{N}\mathrm{d}x\frac{D_{-}D_{+}^2\mathcal{T}}{W}
+\frac{D_{-}}{D_{+}^{2}}\int_{N_{0}}^{N}\mathrm{d}x\frac{D_{+}^{3}\mathcal{T}}{W},\\
\mathcal{X}_4(N)&\equiv\frac{7D_{12,z}}{3D_+^2},\\
\mathcal{X}_5(N)&\equiv\frac{7D_{12,z} \mathcal{X}_{0}}{3D_+^3},\\
\mathcal{X}_6(N)&\equiv e^{-I_1}\int_{N_0}^{N}e^{I_1}\mathcal{T}dx,
\end{align}
while the corresponding scale-dependent kernel functions are given by:
\begin{align}
\hat{\mathcal{K}}_1 &\equiv  \Bigl( \mathcal{K}(k_1) + \mathcal{K}(k_2) \Bigr) (1 - \hat{\mu}_{12}^2),\\
\hat{\mathcal{K}}_2 &\equiv  \mathcal{K}(k)\, (1 - \hat{\mu}_{12}^2),\\
\hat{\mathcal{K}}_3 &\equiv  \mathcal{K}(k) + \Biggl( \frac{k_1}{k_2} \Bigl( \mathcal{K}(k) - \mathcal{K}(k_1) \Bigr) + \frac{k_2}{k_1} \Bigl( \mathcal{K}(k) - \mathcal{K}(k_2) \Bigr) \Biggr) \hat{\mu}_{12}  + \Bigl( \mathcal{K}(k) - \mathcal{K}(k_1) - \mathcal{K}(k_2) \Bigr) \hat{\mu}_{12}^2 ,\\
\hat{\mathcal{K}}_4 &\equiv  \Bigl( \frac{k_1}{k_2} \mathcal{K}(k_1) + \frac{k_2}{k_1} \mathcal{K}(k_2) \Bigr) \hat{\mu}_{12} + \Bigl( \mathcal{K}(k_1) + \mathcal{K}(k_2) \Bigr) \hat{\mu}_{12}^2 ,\\
\hat{\mathcal{K}}_5 &\equiv  \hat{\mu}_{12}^2+\hat{\mu}_{12}\Big(\frac{k_1}{k_2}+\frac{k_2}{k_1}\Big),\\
\hat{\mathcal{K}}_6 &\equiv  (1 - \hat{\mu}_{12}^2)\\
\hat{\mathcal{K}}_7 &\equiv  \hat{\mu}_{12}^2+\frac{1}{2}\hat{\mu}_{12}\Big(\frac{k_1}{k_2}+\frac{k_2}{k_1}\Big).
\end{align}
From Eq. ~\eqref{eq:fk_solution}, the function $f_{kz}$ is then given by
\begin{equation}
    f_{kz}=\mathcal{K}(k)\,\mathcal{X}_6(N).
\end{equation}
This factorization enables the separation of temporal evolution from scale dependence, allowing the complex integrals to be decomposed into pre-computable time-dependent functions $\mathcal{X}_i(N)$ and scale-dependent factors $\mathcal{K}(k)$. While this introduces additional pre-computation steps, it significantly reduces the cost of repeated kernel evaluations. 
The second-order kernels $F_2$ and $G_2$ in Eq. ~\eqref{eq:F_2_G_2_kernel} can be decomposed following the same definition in Eq.~\eqref{eq:F2_hc0} to Eq.~\eqref{eq:G2_hc1}, where \( F_{h_c^0} \) denotes the component of \( F \) at order \( h_c^0 \) and the same as \( G_{h_c^0} \), given by : 
\begin{align}
F_2&=F_{2,h_c^0} +F_{2,h_c^1} , \\
G_2&=G_{2,h_c^0}+ G_{2,h_c^1}  .
\label{eq: F2_G2_Seperation_kernel}
\end{align} %
The zeroth-order contributions are given by:
\begin{align}
F_{2,h_c^0} &= \frac{1}{2}(1+\hat{\mathcal{K}}_5)+\frac{3}{14}\mathcal{X}_4\hat{\mathcal{K}}_6,
\\
G_{2,h_c^0} &=f_z\hat{\mathcal{K}}_7+\frac{3}{14}\Big(2f_z \mathcal{X}_4 + \mathcal{X}_4'\Big)\hat{\mathcal{K}}_6,
\end{align}
while the first-order terms take the form:
\begin{align}
F_{2,h_c^1} &= \sum_{i=1}^{3} \frac{1}{2} \mathcal{X}_i \hat{\mathcal{K}}_i - \frac{3}{14} \mathcal{X}_5 \hat{\mathcal{K}}_1, \label{eq:F2_Separation_kernel} \\
G_{2,h_c^1} 
&= \sum_{i=1}^{3} \left( f_z \mathcal{X}_i + \frac{1}{2}\mathcal{X}_i \mathcal{X}_6 \left( \mathcal{K}_1 + \mathcal{K}_2 \right) + \frac{1}{2} \mathcal{X}_i' \right) \hat{\mathcal{K}}_i 
 - \frac{3}{7} \left( f_z \mathcal{X}_5 + \frac{1}{2} \mathcal{X}_5' \right) \hat{\mathcal{K}}_1 \notag \\
&\quad + \frac{1}{2} \mathcal{X}_6 \hat{\mathcal{K}}_4 
+ \frac{3}{14} \mathcal{X}_4 \mathcal{X}_6 \hat{\mathcal{K}}_1 
- \frac{3}{14} \mathcal{X}_5 \mathcal{X}_6 \left( \mathcal{K}_1 + \mathcal{K}_2 \right) \hat{\mathcal{K}}_1, \label{eq:G2_Separation_kernel}
\end{align}

where $\mathbf{k}=\mathbf{k_1}+\mathbf{k_2}$, $\hat{\mu}_{12}$ is the cosine of the angle between the vectors $\mathbf{k}_1$ and $\mathbf{k}_2$, and $\mathcal{K}_i=\mathcal{K}(k_i)$ . 

\subsection{Third-order kernels}
Following the same decomposition as for the second-order kernels, the scale-time separation can be applied to $\mathcal{I}_{3,z}$ in Eq.~\eqref{eq:LCDM_DA^3_1},
which admits the factorized form:
\begin{align}
\mathcal{I}_{3,z} &=\sum_{i=1}^{6}\tilde{\mathcal{Z}}_{3z,i}  \tilde{\mathcal{K}}_{3z,i}
\end{align}
The time-dependent $\tilde{\mathcal{Z}}_{3z,i}$ and scale-dependent $\tilde{\mathcal{K}}_{3z,i}$ are listed in Table \ref{tab:I3z_decomposition_updated}. Based on this decomposition, the function $D_{123,z}$ in Eq.~\eqref{eq:par_solu_third} can be constructed through:
\begin{equation}
D_{123,z}(k,N)=\sum_{i=1}^{6}\left\{-D_{+}\tilde{\mathcal{K}}_{3z,i}\int_{N_{0}}^{N}\mathrm{d}x\frac{D_{-}\tilde{\mathcal{Z}}_{3z,i}}{W}+D_{-}\tilde{\mathcal{K}}_{3z,i}\int_{N_{0}}^{N}\mathrm{d}x\frac{D_{+}\tilde{\mathcal{Z}}_{3z,i}}{W}\right\}\,.
\end{equation}
\begin{table}
    \centering
    \renewcommand{\arraystretch}{1.5}
    \resizebox{\textwidth}{!}{%
    \begin{tabular}{|c|c|c|}    
    \hline
    index $i$ & Time-dependent coefficients $\tilde{\mathcal{Z}}_{3z,i}$& Scale-dependent terms $\tilde{\mathcal{K}}_{3z,i}$\\
    \hline
    1 & $D_+^3(\mathcal{F} f_z + f_z')$& $\left\{\alpha_{1,23}(1+\hat{\mathcal{K}}_5)+2\alpha_{23,1}\hat{\mathcal{K}}_7\right\}_{\text{cyc}}$\\
    \hline
    2 & $D_+^3 f_z^2$ & $\left\{3\alpha_{1,23}(1+\hat{\mathcal{K}}_5)+(6\alpha_{23,1}+ 4\beta_{1,23})\hat{\mathcal{K}}_7\right\}_{\text{cyc}}$\\
    \hline
    3 & $D_+^3 (\mathcal{F} f_z + f_z') \mathcal{X}_4$& $\left\{\frac{3}{7}(\alpha_{1,23}+2\alpha_{23,1})\hat{\mathcal{K}}_6\right\}_{\text{cyc}}$\\
    \hline
    4 & $D_+^3 f_z^2 \mathcal{X}_4$& $\left\{\frac{3}{7}(3\alpha_{1,23}+6\alpha_{23,1}+4\beta_{1,23})\hat{\mathcal{K}}_6\right\}_{\text{cyc}}$\\
    \hline
    5 & $D_+^3 (\mathcal{X}_4''+\mathcal{F} \mathcal{X}_4')$& $\left\{\frac{3}{7}\alpha_{23,1}\hat{\mathcal{K}}_6\right\}_{\text{cyc}}$\\
    \hline
    6 & $D_+^3 f_z \mathcal{X}_4'$& $\left\{\frac{3}{7}(\alpha_{1,23}+5\alpha_{23,1}+2\beta_{1,23})\hat{\mathcal{K}}_6\right\}_{\text{cyc}}$\\
    \hline
    \end{tabular}
    }
    \caption{Summary of the six terms contributing to $\mathcal{I}_{3,z}$, with time- and scale-dependent components separated using kernel functions $\hat{\mathcal{K}}_5$, $\hat{\mathcal{K}}_6$, and $\hat{\mathcal{K}}_7$.}
    \label{tab:I3z_decomposition_updated}
\end{table}
Similarly, this separation can be extended to $R$ in Eq. ~\eqref{eq:dd_D123} as:
\begin{equation}
\quad R = \sum_{i=1}^{23}\tilde{\mathcal{Z}}_{i} \cdot \tilde{\mathcal{K}}_i,
\end{equation} %
with the components $\tilde{\mathcal{Z}}_{i}$ and $\tilde{\mathcal{K}}_i$ detailed in Table \ref{tab:R_terms}.
Combining these decompositions above, the complete source term $\hat{\mathcal{I}}_{3}$  takes the form:
\begin{equation}
\hat{\mathcal{I}}_{3}=6\left(3D_z+\mathcal{X}_{0}\left(\mathcal{K}_1+\mathcal{K}_2+\mathcal{K}_3\right)\right) \sum_{i=1}^{23}\tilde{\mathcal{Z}}_{i} \cdot \tilde{\mathcal{K}}_i-\mathcal{I}_{3,z}+\mathcal{T}D_{123,z}\mathcal{K}.
\end{equation}
The $\mathcal{A}_3$ is then constructed as:
\begin{align}
    \mathcal{A}_3 &= \frac{7D_{123,z}}{3D_{+}^{3}}\left[1-\left(\mathcal{K}_1+\mathcal{K}_2+\mathcal{K}_3\right)\frac{\mathcal{X}_{0}}{D_+}\right] \notag \\
    &\quad + \frac{7}{3}\left\{ -\frac{1}{D_{+}}\int_{N_{0}}^{N}\mathrm{d}x\frac{D_{-}\hat{\mathcal{I}}_{3}}{W}+\frac{D_{-}}{D_{+}^{2}}\int_{N_{0}}^{N}\mathrm{d}x\frac{D_{+}\hat{\mathcal{I}}_{3}}{W}\right\}, 
\end{align}
while the coefficient is given by:
\begin{equation} \label{eq:hat_alpha}
\hat{\alpha}(\mathbf{k}_1, \mathbf{k}_2, \mathbf{k}_3)=\Big\{\alpha_{1,23}(f_z+\mathcal{X}_6\mathcal{K}_1)F_{2}(\mathbf{k}_{2},\mathbf{k}_{3})+\alpha_{13,2}G_{2}(\mathbf{k}_{1},\mathbf{k}_{3})\Big\}_{\text{cyc}}.
\end{equation}
Finally, the third-order kernels are obtained as:
\begin{align}
F_3 &= \frac{1}{14}\mathcal{A}_3 \,,   \\
G_3 &= \frac{1}{14}\mathcal{A}_3' + \frac{1}{14}\mathcal{A}_3 
\left(3f_z+\mathcal{X}_6(\mathcal{K}_1+\mathcal{K}_2+\mathcal{K}_3) \right) - \frac{1}{3} \hat{\alpha}\,. 
\end{align}

\begin{table}
    \centering
    \resizebox{\textwidth}{!}{
    \begin{tabular}{|c|c|c|}
        \hline
        index $i$& Time-dependent coefficients $\mathcal{Z}_{kz,i}$& Scale-dependent terms $\mathcal{K}_{kz,i}$\\
        \hline
        1 & $\mathcal{F} f_z+3f_z^2+f_z'$& $\left\{\frac{1}{3}\alpha_{23,1} \hat{\mathcal{K}}_7\right\}_{\mathrm{cyc}}$\\
        \hline
        2& $2(\mathcal{F} f_z + 3f_z^2 + f_z') \mathcal{X}_j + (\mathcal{F} + 5f_z) \mathcal{X}_j' + \mathcal{X}_j''$& $\left\{\frac{1}{6}\alpha_{23,1} \hat{\mathcal{K}}_j\right\}_{\mathrm{cyc}}$\\
        \hline
        3& $2(\mathcal{F} f_z + 3f_z^2 + f_z') \mathcal{X}_4 + (\mathcal{F} + 5f_z) \mathcal{X}_4' + \mathcal{X}_4''$& $\left\{\frac{1}{14}\alpha_{23,1} \hat{\mathcal{K}}_6\right\}_{\mathrm{cyc}}$\\
        \hline
        4& $2(\mathcal{F} f_z + 3f_z^2 + f_z') \mathcal{X}_5 + (\mathcal{F} + 5f_z) \mathcal{X}_5' + \mathcal{X}_5''$& $\left\{-\frac{1}{14}\alpha_{23,1} \hat{\mathcal{K}}_1\right\}_{\mathrm{cyc}}$\\
        \hline
        5& $\mathcal{F}\mathcal{X}_6$& $\left\{\frac{1}{6}\left(\alpha_{1,23}\mathcal{K}_1+\alpha_{23,1}\hat{\mathcal{K}}_4 +\alpha_{1,23}\mathcal{K}_1\hat{\mathcal{K}}_5\right)\right\}_{\mathrm{cyc}}$\\
        \hline
        6& $f_z \mathcal{X}_6$& $\begin{gathered}
        \left\{\frac{1}{2}\left(\alpha_{1,23}\mathcal{K}_1+\alpha_{23,1}\hat{\mathcal{K}}_4 +\alpha_{1,23}\mathcal{K}_1\hat{\mathcal{K}}_5\right)\right.\\
        \left.+\frac{1}{3}\left(\alpha_{23,1}\hat{\mathcal{K}}_7(\mathcal{K}_1+\mathcal{K}_2+\mathcal{K}_3) +2\beta_{1,23}\mathcal{K}_1\hat{\mathcal{K}}_7\right)\right\}_{\mathrm{cyc}}
        \end{gathered}$\\
        \hline
        7& $\mathcal{F}\mathcal{X}_6\mathcal{X}_j$& $\left\{\frac{1}{6}\hat{\mathcal{K}}_j\left[\mathcal{K}_1(\alpha_{1,23}+\alpha_{23,1})+\mathcal{K}_2\alpha_{23,1}\right]\right\}_{\mathrm{cyc}}$\\
        \hline
        8& $f_z \mathcal{X}_6\mathcal{X}_j$& $\begin{gathered}
        \left\{\frac{1}{6}\hat{\mathcal{K}}_j\left[(3\alpha_{1,23}+5\alpha_{23,1}+4\beta_{1,23})\mathcal{K}_1\right.\right.
        \left.\left.+5\alpha_{23,1}\mathcal{K}_2+2\alpha_{23,1}\mathcal{K}_3\right]\right\}_{\mathrm{cyc}}
        \end{gathered}$\\
        \hline
        9& $\mathcal{X}_6\mathcal{X}_j'$& $\begin{gathered}
        \left\{\frac{1}{6}\hat{\mathcal{K}}_j\left[(\alpha_{1,23}+2\alpha_{23,1}+2\beta_{1,23})\mathcal{K}_1\right.\right.
        \left.\left.+2\alpha_{23,1}\mathcal{K}_2+\alpha_{23,1}\mathcal{K}_3\right]\right\}_{\mathrm{cyc}}
        \end{gathered}$\\
        \hline
        10& $\mathcal{F} \mathcal{X}_4 \mathcal{X}_6$& $\left\{\frac{1}{14}\left[\alpha_{23,1} \hat{\mathcal{K}}_1 + \alpha_{1,23} \mathcal{K}_1 \hat{\mathcal{K}}_6\right]\right\}_{\mathrm{cyc}}$\\
        \hline
        11& $f_z \mathcal{X}_4 \mathcal{X}_6$ & $\begin{gathered}
        \left\{\frac{1}{14}\left[3\alpha_{23,1} \hat{\mathcal{K}}_1 + \hat{\mathcal{K}}_6\left((3\alpha_{1,23} + 2\alpha_{23,1} + 4\beta_{1,23}) \mathcal{K}_1\right.\right.\right.\\
        \left.\left.\left. + 2\alpha_{23,1}(\mathcal{K}_2 + \mathcal{K}_3)\right)\right]\right\}_{\mathrm{cyc}}
        \end{gathered}$\\
        \hline
        12& $\mathcal{X}_4' \mathcal{X}_6$ & $\begin{gathered}
        \left\{\frac{1}{14}\left[\alpha_{23,1} \hat{\mathcal{K}}_1 + \hat{\mathcal{K}}_6\left((\alpha_{1,23} + \alpha_{23,1} + 2\beta_{1,23}) \mathcal{K}_1\right.\right.\right.\\
        \left.\left.\left. + \alpha_{23,1}(\mathcal{K}_2 + \mathcal{K}_3)\right)\right]\right\}_{\mathrm{cyc}}
        \end{gathered}$\\
        \hline
        13& $\mathcal{F} \mathcal{X}_5 \mathcal{X}_6$& $\left\{-\frac{\hat{\mathcal{K}}_1}{14}\left[\alpha_{1,23} \mathcal{K}_1 + \alpha_{23,1}(\mathcal{K}_1 + \mathcal{K}_2)\right]\right\}_{\mathrm{cyc}}$\\
        \hline
        14& $f_z \mathcal{X}_5 \mathcal{X}_6$ & $\begin{gathered}
        \left\{-\frac{\hat{\mathcal{K}}_1}{14}\left[3\alpha_{1,23} \mathcal{K}_1 + \alpha_{23,1}(5\mathcal{K}_1 + 5\mathcal{K}_2 + 2\mathcal{K}_3)\right.\right.\\
        \left.\left. + 4\beta_{1,23} \mathcal{K}_1\right]\right\}_{\mathrm{cyc}}
        \end{gathered}$\\
        \hline
        15& $\mathcal{X}_5' \mathcal{X}_6$ & $\begin{gathered}
        \left\{-\frac{\hat{\mathcal{K}}_1}{14}\left[\alpha_{1,23} \mathcal{K}_1 + \alpha_{23,1}(2\mathcal{K}_1 + 2\mathcal{K}_2 + \mathcal{K}_3)\right.\right.\\
        \left.\left. + 2\beta_{1,23} \mathcal{K}_1\right]\right\}_{\mathrm{cyc}}
        \end{gathered}$\\
        \hline
        16& $\mathcal{X}_6^2$ & $\begin{gathered}
        \left\{ \frac{\alpha_{1,23}}{6} \mathcal{K}_1 (\mathcal{K}_1 + \mathcal{K}_2 + \mathcal{K}_3)(1 + \hat{\mathcal{K}}_5)\right.\\
        \left. + \frac{\alpha_{23,1}}{6} \hat{\mathcal{K}}_4 (\mathcal{K}_1 + \mathcal{K}_2 + \mathcal{K}_3) + \frac{\beta_{1,23}}{3} \mathcal{K}_1 \hat{\mathcal{K}}_4 \right\}_{\mathrm{cyc}}
        \end{gathered}$\\
        \hline
        17& $\mathcal{X}_j \mathcal{X}_6^2$& $\begin{gathered}
        \left\{\frac{\hat{\mathcal{K}}_j}{6}\left[\alpha_{1,23} \mathcal{K}_1(\mathcal{K}_1 + \mathcal{K}_2 + \mathcal{K}_3) + \alpha_{23,1}(\mathcal{K}_1^2 + 2\mathcal{K}_1 \mathcal{K}_2 + \mathcal{K}_2^2\right.\right.
   \\ \left.\left. + \mathcal{K}_1 \mathcal{K}_3 + \mathcal{K}_2 \mathcal{K}_3) + 2\beta_{1,23}\hat{ \mathcal{K}}_1(\mathcal{K}_1 + \mathcal{K}_2)\right]\right\}_{\mathrm{cyc}}
        \end{gathered}$\\
        \hline
        18& $\mathcal{X}_4 \mathcal{X}_6^2$ & $\begin{gathered}
        \left\{\frac{1}{14}\left[\hat{\mathcal{K}}_1(\alpha_{23,1}(\mathcal{K}_1 + \mathcal{K}_2 + \mathcal{K}_3) + 2\beta_{1,23} \mathcal{K}_1)\right.\right.\\
        \left.\left. + \alpha_{1,23} \mathcal{K}_1 \hat{\mathcal{K}}_6(\mathcal{K}_1 + \mathcal{K}_2 + \mathcal{K}_3)\right]\right\}_{\mathrm{cyc}}
        \end{gathered}$\\
        \hline
        19& $\mathcal{X}_5 \mathcal{X}_6^2$ & $\begin{gathered}
        \left\{-\frac{\hat{\mathcal{K}}_1}{14}\left[\alpha_{1,23} \mathcal{K}_1(\mathcal{K}_1 + \mathcal{K}_2 + \mathcal{K}_3) + \alpha_{23,1}(\mathcal{K}_1^2 + 2\mathcal{K}_1 \mathcal{K}_2\right.\right.\\
        \left.\left. + \mathcal{K}_2^2 + \mathcal{K}_1 \mathcal{K}_3 + \mathcal{K}_2 \mathcal{K}_3) + 2\beta_{1,23} \mathcal{K}_1(\mathcal{K}_1 + \mathcal{K}_2)\right]\right\}_{\mathrm{cyc}}
        \end{gathered}$\\
        \hline
        20& $\mathcal{X}_6'$ & $\left\{\frac{1}{6}(\alpha_{1,23} \mathcal{K}_1+\alpha_{23,1} \hat{\mathcal{K}}_4+\alpha_{1,23} \mathcal{K}_1 \hat{\mathcal{K}}_5)\right\}_{\mathrm{cyc}}$\\
        \hline
        21& $\mathcal{X}_j \mathcal{X}_6'$& $\left\{ \frac{1}{6} \hat{\mathcal{K}}_j \left[ (\alpha_{1,23} + \alpha_{23,1}) \mathcal{K}_1 + \alpha_{23,1} \mathcal{K}_2 \right] \right\}_{\mathrm{cyc}}$\\
        \hline
        22& $\mathcal{X}_4 \mathcal{X}_6'$ & $\left\{ \frac{1}{14} \left( \alpha_{23,1} \hat{\mathcal{K}}_1 + \alpha_{1,23} \mathcal{K}_1 \hat{\mathcal{K}}_6 \right) \right\}_{\mathrm{cyc}}$\\
        \hline
        23& $\mathcal{X}_5 \mathcal{X}_6'$ & $\left\{-\frac{1}{14}\hat{\mathcal{K}}_1\left((\alpha_{1,23}+\alpha_{23,1})\mathcal{K}_1+ \alpha_{23,1}\mathcal{K}_2\right)\right\}_{\mathrm{cyc}}$\\
    \hline
\end{tabular}
}
   \caption{Summary of the terms contributing to $R$, with time- and scale-dependent components separated. Index $j$ represents the cyclic indices $j = 1, 2, 3$. }
   \label{tab:R_terms}
\end{table}
\section{Summary of results}    
\label{sec:summary}

For convenience, we collect here the kernel equations. We recall that
the background and linear functions $\mathcal{F},S_{kz},D_{\pm},D_{kz},f$
are all defined in Sec. \ref{sec:Linear}.

The second-order kernels are
\begin{align}
F_{2}(\mathbf{k}_{1},\mathbf{k}_{2}) & =\frac{1}{2}+\frac{3}{14}\mathcal{A}+\left(\frac{1}{2}-\frac{3}{14}\mathcal{B}\right)\frac{(\mathbf{k}_{1}\cdot\mathbf{k}_{2})^{2}}{k_{1}^{2}k_{2}^{2}}+\frac{\mathbf{k}_{1}\cdot\mathbf{k}_{2}}{2k_{1}k_{2}}\left(\frac{k_{2}}{k_{1}}+\frac{k_{1}}{k_{2}}\right)\,,\label{eq:F2kernel-1}\\
G_{2}(\mathbf{k}_{1},\mathbf{k}_{2}) & =\frac{3\mathcal{A}(f_{1}+f_{2})+3\mathcal{A}'}{14}+\left(\frac{f_{1}+f_{2}}{2}-\frac{3\mathcal{B}(f_{1}+f_{2})+3\mathcal{B}'}{14}\right)\frac{(\mathbf{k}_{1}\cdot\mathbf{k}_{2})^{2}}{k_{1}^{2}k_{2}^{2}}\nonumber\\&+\frac{\mathbf{k}_{1}\cdot\mathbf{k}_{2}}{2k_{1}k_{2}}\left(\frac{f_{2}k_{2}}{k_{1}}+\frac{f_{1}k_{1}}{k_{2}}\right)\,,\label{eq:F_2_G_2_kernel-1}
\end{align}
with 
\begin{align}
\mathcal{A} & =\frac{7D_{12,z}}{3D_{+}^{2}}\left[1+\int_{N_{0}}^{N}\mathrm{d}x\frac{D_{-}D_{+}\big(S_{kz}(k_{1})+S_{kz}(k_{2})\big)}{W}-\frac{D_{-}}{D_{+}}\int_{N_{0}}^{N}\mathrm{d}x\frac{D_{+}^{2}\big(S_{kz}(k_{1})+S_{kz}(k_{2})\big)}{W}\right]+\nonumber \\
 & \frac{7}{3}\left\{ -\frac{1}{D_{+}}\int_{N_{0}}^{N}\mathrm{d}x\frac{D_{-}\hat{\mathcal{I}}_{\mathcal{A}}}{W}+\frac{D_{-}}{D_{+}^{2}}\int_{N_{0}}^{N}\mathrm{d}x\frac{D_{+}\hat{\mathcal{I}}_{\mathcal{A}}}{W}\right\} \,,\label{eq:Horndeski_a-1}\\
\mathcal{B} & =\frac{7D_{12,z}}{3D_{+}^{2}}\left[1+\int_{N_{0}}^{N}\mathrm{d}x\frac{D_{-}D_{+}\big(S_{kz}(k_{1})+S_{kz}(k_{2})\big)}{W}-\frac{D_{-}}{D_{+}}\int_{N_{0}}^{N}\mathrm{d}x\frac{D_{+}^{2}\big(S_{kz}(k_{1})+S_{kz}(k_{2})\big)}{W}\right]+\\
 & \frac{7}{3}\left\{ -\frac{1}{D_{+}}\int_{N_{0}}^{N}\mathrm{d}x\frac{D_{-}\hat{\mathcal{I}}_{\mathcal{B}}}{W}+\frac{D_{-}}{D_{+}^{2}}\int_{N_{0}}^{N}\mathrm{d}x\frac{D_{+}\hat{\mathcal{I}}_{\mathcal{B}}}{W}\right\} \,,
\end{align}
where 
\begin{align}
\hat{\mathcal{I}}_{\mathcal{A}}&\equiv S_{z}D_{z}(D_{kz}(k_{1})+D_{kz}(k_{2}))\\&+ \left[S_{kz}(k)+(S_{kz}(k)-S_{kz}(k_{1}))\frac{\mathbf{k}_{1}\cdot\mathbf{k}_{2}}{k_{2}^{2}}+(S_{kz}(k)-S_{kz}(k_{2}))\frac{\mathbf{k}_{1}\cdot\mathbf{k}_{2}}{k_{1}^{2}}\right]D_{z}^{2}+S_{kz}(k)D_{12,z}\\\hat{\mathcal{I}}_{\mathcal{B}}&\equiv S_{z}D_{z}(D_{kz}(k_{1})+D_{kz}(k_{2}))+\Biggl[S_{kz}(k_{1})+S_{kz}(k_{2})-S_{kz}(k)\Biggr]D_{z}^{2}+S_{kz}(k)D_{12,z}.
\end{align}
and
\begin{align}
D_{12,z}(N) & \frac{3}{2}\left[-D_{+}(N)\int_{N_{0}}^{N}\mathrm{d}x\,\frac{\Omega_{m}(x)D_{-}(x)D_{+}^{2}(x)h_{1}(x)}{W(D_{+},D_{-})}+D_{-}(N)\int_{N_{0}}^{N}\mathrm{d}x\,\frac{\Omega_{m}(x)D_{+}^{3}(x)h_{1}(x)}{W(D_{+},D_{-})}\right]\,.
\end{align}
The third-order kernels are
\begin{align}
F_{3} & =\frac{1}{14}\mathcal{A}_{3}\,,\label{eq:final_F3-1}\\
G_{3} & =\frac{1}{14}\mathcal{A}_{3}'+\frac{1}{14}\mathcal{A}_{3}\left(f_{1}+f_{2}+f_{3}\right)-\frac{1}{3}\hat{\alpha}\,,\label{eq:final_G3-1}
\end{align}
where
\begin{align}
\mathcal{A}_{3}&=\frac{7D_{123,z}}{3D_{+}^{3}}\left[1+\int_{N_{0}}^{N}\mathrm{d}x\frac{D_{-}D_{+}\big(S_{kz}(k_{1})+S_{kz}(k_{2})+S_{kz}(k_{3})\big)}{W}\right.\nonumber\\&\left.-\frac{D_{-}}{D_{+}}\int_{N_{0}}^{N}\mathrm{d}x\frac{D_{+}^{2}\big(S_{kz}(k_{1})+S_{kz}(k_{2})+S_{kz}(k_{3})\big)}{W}\right]+\nonumber\\&\frac{7}{3}\left\{ -\frac{1}{D_{+}}\int_{N_{0}}^{N}\mathrm{d}x\frac{D_{-}\hat{\mathcal{I}}_{3}}{W}+\frac{D_{-}}{D_{+}^{2}}\int_{N_{0}}^{N}\mathrm{d}x\frac{D_{+}\hat{\mathcal{I}}_{3}}{W}\right\} \,,\label{eq:Horndeski_{3}a-1}
\end{align}
with 
\begin{align}
\hat{\mathcal{I}}_{3} & =\Big(2\hat{\alpha}'_{h_{c}^{0}}+4\hat{\beta}_{h_{c}^{0}}+2\hat{\alpha}_{h_{c}^{0}}(3f_{z}+\mathcal{F})\Big)\Big[D_{kz}(k_{1})+D_{kz}(k_{2})+D_{kz}(k_{3})\Big]D_{z}^{2}+2\hat{\alpha}'_{h_{c}^{1}}D_{z}^{3}+4\hat{\beta}_{h_{c}^{1}}D_{z}^{3}\nonumber \\
 & +2\hat{\alpha}_{h_{c}^{1}}(3f_{z}+\mathcal{F})D_{z}^{3}+2\hat{\alpha}_{h_{c}^{0}}\Big[f_{kz}(k_{1})+f_{kz}(k_{2})+f_{kz}(k_{3})\Big]D_{z}^{3}+\frac{3}{2}\Omega_{m}h_{c}D_{123,{z}}\,,\label{eq:i3a-1-1}
\end{align}
(explicit expressions for $\hat{\alpha}{}_{h_{c}^{0}},\hat{\alpha}_{h_{c}^{1}},\hat{\beta}_{h_{c}^{0}},\hat{\beta}_{h_{c}^{1}}$
are provided in App.~\ref{sec:detailed}). Moreover
\begin{equation}
\hat{\alpha}(\mathbf{k}_{1},\mathbf{k}_{2},\mathbf{k}_{3})=\Big\{\alpha_{1,23}f_{1}F_{2}(\mathbf{k}_{2},\mathbf{k}_{3})+\alpha_{13,2}G_{2}(\mathbf{k}_{1},\mathbf{k}_{3})\Big\}_{\text{cyc}}\,,\label{eq: hat_a-1}
\end{equation}
and
\begin{equation}
D_{123,kz}(\textbf{k}_{1},\textbf{k}_{2},\textbf{k}_{3},N)=-D_{+}(N)\int_{N_{0}}^{N}\mathrm{d}x\frac{D_{-}(x)\hat{\mathcal{I}}_{3}(x)}{W\left(D_{+}(x),D_{-}(x)\right)}+D_{-}(N)\int_{N_{0}}^{N}\mathrm{d}x\frac{D_{+}(x)\hat{\mathcal{I}}_{3}(x)}{W\left(D_{+}(x),D_{-}(x)\right)}\,.\label{eq:par_solu_third-1}
\end{equation}

\bibliographystyle{ieeetr} 
\clearpage
\bibliography{kernel_paper,references,literature,references2}
\end{document}